\documentclass[sigconf]{acmart}

\usepackage[utf8]{inputenc} % allow utf-8 input
\usepackage[T1]{fontenc}    % use 8-bit T1 fonts
\usepackage{hyperref}       % hyperlinks
\usepackage{url}            % simple URL typesetting
\usepackage{booktabs}       % professional-quality tables
\usepackage{amsfonts}       % blackboard math symbols
\usepackage{nicefrac}       % compact symbols for 1/2, etc.
\usepackage{microtype}      % microtypography
\usepackage{xcolor}         % colors
\usepackage{multirow}       % table
\usepackage{algorithm,algpseudocode}
\usepackage{graphicx}
\usepackage{cite}
\usepackage{amsmath}
\usepackage{amsthm}
\newtheorem{theorem}{Theorem}
\newtheorem{corollary}{Corollary}
\newtheorem{lemma}{Lemma}
\newtheorem{definition}{Definition}
\newtheorem{assumption}{Assumption}
\usepackage{appendix}
\usepackage{subfigure}
\usepackage{bbm}
\usepackage{bm}
\usepackage{mathrsfs}
\usepackage{enumitem}
\usepackage{mathtools} % amsmath with fixes and additions
\usepackage{tikz} 
\usepackage{fancyhdr}
\AtBeginDocument{%
  \providecommand\BibTeX{{%
    \normalfont B\kern-0.5em{\scshape i\kern-0.25em b}\kern-0.8em\TeX}}}

\copyrightyear{2023}
\acmYear{2023}
\setcopyright{acmlicensed}\acmConference[CIKM '23]{Proceedings of the 32nd ACM International Conference on Information and Knowledge Management}{October 21--25, 2023}{Birmingham, United Kingdom}
\acmBooktitle{Proceedings of the 32nd ACM International Conference on Information and Knowledge Management (CIKM '23), October 21--25, 2023, Birmingham, United Kingdom}
\acmPrice{15.00}
\acmDOI{10.1145/3583780.3615135}
\acmISBN{979-8-4007-0124-5/23/10}

\begin{document}

\title{Dynamic Embedding Size Search with Minimum Regret for Streaming Recommender System}
% Minimizing Embedding Size Regret for Streaming Recommender Systems

%\author{Anonymous authors}

\author{Bowei He}
%\email{webmaster@marysville-ohio.com}
\affiliation{%
  \institution{City University of Hong Kong}
  %\streetaddress{P.O. Box 1212}
  %\city{Hong Kong}
  \country{Hong Kong SAR}
}
\email{boweihe2-c@my.cityu.edu.hk}

\author{Xu He}
\affiliation{%
  \institution{Huawei Noah's Ark Lab}
  %\streetaddress{1 Th{\o}rv{\"a}ld Circle}
  \city{Shenzhen}
  \country{China}
  }
\email{hexu27@huawei.com}

\author{Renrui Zhang}
\affiliation{%
  \institution{The Chinese University of Hong Kong}
  \country{Hong Kong SAR}
}
\email{zhangrenrui@pjlab.org.cn	}

\author{Yingxue Zhang}
\affiliation{%
  \institution{Huawei Noah's Ark Lab Montreal}
  \city{Montreal}
  \country{Canada}
}
\email{yingxue.zhang@huawei.com}

\author{Ruiming Tang}
\affiliation{%
 \institution{Huawei Noah's Ark Lab}
 %\streetaddress{Rono-Hills}
 \city{Shenzhen}
 %\state{Arunachal Pradesh}
 \country{China}
 }
 \email{tangruiming@huawei.com}

\author{Chen Ma}
\authornote{Corresponding author}
\affiliation{%
  \institution{City University of Hong Kong}
 % \streetaddress{30 Shuangqing Rd}
  %\city{Hong Kong SAR}
  %\state{Beijing Shi}
  \country{Hong Kong SAR}}
\email{chenma@cityu.edu.hk}

\renewcommand{\shortauthors}{Bowei He et al.}


\begin{abstract}
With the continuous increase of users and items, conventional recommender systems trained on static datasets can hardly adapt to changing environments. The high-throughput data requires the model to be updated in a timely manner for capturing the user interest dynamics, which leads to the emergence of streaming recommender systems. Due to the prevalence of deep learning-based recommender systems, the embedding layer is widely adopted to represent the characteristics of users, items, and other features in low-dimensional vectors. However, it has been proved that setting an identical and static embedding size is sub-optimal in terms of recommendation performance and memory cost, especially for streaming recommendations. To tackle this problem, we first rethink the streaming model update process and model the dynamic embedding size search as a bandit problem. Then, we analyze and quantify the factors that influence the optimal embedding sizes from the statistics perspective. Based on this, we propose the \textbf{D}ynamic \textbf{E}mbedding \textbf{S}ize \textbf{S}earch (\textbf{DESS}) method to minimize the embedding size selection regret on both user and item sides in a non-stationary manner. Theoretically, we obtain a sublinear regret upper bound superior to previous methods. Empirical results across two recommendation tasks on four public datasets also demonstrate that our approach can achieve better streaming recommendation performance with lower memory cost and higher time efficiency. 
\end{abstract}


\ccsdesc[500]{Information systems~Recommender systems}
%\ccsdesc[500]{Computing methodologies~Online learning settings}
% %\ccsdesc[300]{Computer systems organization~Redundancy}
% \ccsdesc{Computing methodologies~Neural networks}

\keywords{Streaming recommender system; Embedding size search; Contextual bandit; Automated machine learning}


\maketitle

\section{Introduction}
% recommender system -> streaming RS -> embedding (embedding size) is important in RS -> current methods in embedding size search -> What is the problem of current methods in streaming RS -> To tackle these problems, we propose ... -> contribution summary
Recommender systems (RS) have been widely adopted to reduce information overload and satisfy users' diverse needs. Considering the ever-increasing users and items, as well as users' continuous interest shift, conventional static RS, however, can hardly adapt to the evolving environment. To tackle such challenges, streaming recommender systems (SRS), whose recommendation strategy updates in a dynamic manner, was proposed in the last decade along with the rapid accumulation of massive data from online applications like Google and Twitter~\citep{das2007google, song2008real,chen2013terec, chandramouli2011streamrec, diaz2012real, he2023dynamically}. %Early stage works mainly focus on promoting the traditional static recommender algorithms like collaborative filtering and matrix factorization to the online fashion~\citep{chandramouli2011streamrec,rendle2008online, devooght2015dynamic}. The data streams processed by such models are sophisticatedly designed. Recent researches in streaming recommender systems pay much attention on detecting the data distribution shift and capturing the dynamics. Some methods~\citep{rendle2012bpr, wang2018streaming} utilize bayesian inference technique to tackle the users' interest drifts and model new users or items in the large-volume and high-velocity data. Sequential prediction models like RNN are also adopted to model the data stream as a data tuple sequence~\citep{hidasi2015session, wu2017recurrent, beutel2018latent}. 
Recently, deep learning-based recommender systems~\citep{zhang2016collaborative, he2017neural} make a breakthrough in improving the recommendation performance, which provides a new direction for the evolution of streaming recommender systems.

% Embedding structure is widely used in the recommender system thanks to its innate property converting the discrete items to the continuous vectors in recommender systems~\citep{zhang2016collaborative, he2017neural, wang2019neural, sun2021hgcf}. Owing to the inherent differences between various items and users, conventional design assigning uniform embedding sizes for each ID restricts the model performance improvement and brings huge memory cost. To solve this problem, embedding size search is first proposed in~\citep{joglekar2020neural} accompanied by the rapid development of data driven automated neural network structure design (NAS) in representation learning on the computer vision and natual language process tasks~\citep{liu2018darts, xie2018snas, brock2017smash, pham2018efficient, luo2018neural}. Recently, embedding size search has received widespread attention~\citep{joglekar2020neural, liu2021learnable, kang2020learning, ginart2021mixed} and various methods have been proposed to search for ID-specific embedding sizes. Most of them adopt an external controller or the differential search to select the embedding sizes from a discrete or continuous candidate space\citep{joglekar2020neural, liu2020automated, zhao2021autodim, cheng2020differentiable}.  Due to the large-scale application of streaming recommendation systems, dynamic embedding size search problem has also been gradually investigated~\citep{liu2020automated, zhao2021autodim, veloso2021hyper,zhao2021autodim}.

To enable the success of deep RS, embedding learning plays a central role in representing users, items, and other features in low-dimensional vectors. Due to the inherent characteristics of different users and items, the conventional design that assigns an identical embedding size to each user or item limits the model performance and brings huge memory costs. To solve this problem, embedding size search is introduced in~\citep{joglekar2020neural} accompanied by the rapid development of the automated neural network structure design in computer vision and natural language processing tasks~\citep{liu2018darts, xie2018snas, brock2017smash, pham2018efficient, luo2018neural}. Recently, embedding size search has received widespread attention~\citep{joglekar2020neural, liu2021learnable, kang2020learning, ginart2021mixed}, and various methods have been proposed to search for ID-specific embedding sizes. Most of them adopt an external controller or the differential search to decide the embedding sizes from a discrete or continuous candidate space~\citep{joglekar2020neural, liu2020automated, zhao2021autodim, cheng2020differentiable, siyi2021iclr, lyu2022optembed}.  Due to the broad application of streaming recommender systems, it has also been noticed that assigning a constant embedding size for users or items along the timeline will lead to unsatisfactory performance. Correspondingly, the dynamic embedding size search problem has also been gradually investigated and several methods~\citep{zhaok2021autoemb, liu2018darts,liu2020automated, veloso2021hyper} have been proposed to search the best time-varying embedding sizes.

Although the aforementioned approaches are effective and insightful, there are still several avenues for improvement. First, previous embedding size search methods~\citep{liu2018darts,liu2020automated, zhao2021autodim, veloso2021hyper} with neural network-based or reinforcement learning-based controllers require large amounts of interaction data to converge, which drags down their effectiveness in the early stages of the recommendation stream. Moreover, the neural network-based method can hardly balance the exploration and exploitation during the online embedding size search. Second, existing methods~\citep{liu2018darts,joglekar2020neural, liu2020automated, zhao2021autodim, veloso2021hyper} only consider the browsing frequency throughout the whole historical record when deciding the appropriate embedding sizes. Nevertheless, this is far from reflecting users' recent browsing behavior characteristics. In fact, the optimal embedding size is mainly associated with the information amount of users' browsing records, which should be determined by both the frequency and diversity of browsing records. Third, in previous works~\citep{liu2018darts,liu2020automated, veloso2021hyper,zhao2021autodim}, the model update process still follows the conventional update pattern of static models, which optimizes the model on the training set until convergence and then evaluating it on the untimely test set. However, in a real SRS like Twitter, the system needs to recommend content to users according to their real-time interests when they post tweets, which means training and testing should alternate over time. Therefore, the optimization objective of dynamic embedding size search should consider the model's performance at each timestep throughout the whole timeline. Moreover, existing methods often suffer from huge memory costs and low time efficiency which make it difficult to put them into practical use.
%\begin{figure}[ht]
%    \centering
%    \includegraphics[width=0.47\textwidth]{Figures/sizes_performance_new.png}
%    \caption{The recommendation accuracy of models with different embedding sizes along the data stream of Amazon.
%    }
%    \label{fig:SRS}
%    \vspace{-0.3cm}
%\end{figure}

To address these issues, we rethink the streaming recommendation model update process, and first model the dynamic optimal embedding size search as a cumulative regret minimization bandit problem. Then, we justify the change of the embedding size by analyzing and quantifying the user behavior characteristics via two indicators proposed in this paper. On this basis, we propose the non-stationary LinUCB bandit-based \textbf{D}ynamic \textbf{E}mbedding \textbf{S}ize \textbf{S}earch (\textbf{DESS}) algorithm to minimize the performance regret. In the method design, the weighted forgetting mechanism is adopted to reduce interference from outdated data and help the search policy pay more attention to recent user behaviors. We provide a sublinear dynamic regret upper bound which guarantees the effectiveness of our method. To help validate the superiorities of our approach, we design an embedding size adaptive neural network based on Neural Collaborative Filtering~\citep{wang2019neural}, whose \textit{embedding input} part can be shared among different base recommendation models. Following previous works~\citep{liu2018darts, zhaok2021autoemb, liu2020automated}, we conduct the top-$k$ recommendation and rating score prediction tasks on four public recommendation datasets. The experimental results demonstrate the effectiveness of our method with lower memory cost and higher time efficiency.
\par To summarize, the main contributions of this work are as follows:
%\vspace{-0.2cm}
\begin{itemize}[leftmargin=*]
    \item  We model the dynamic embedding size search as the cumulative regret minimization bandit and propose a non-stationary LinUCB-based algorithm \textbf{DESS} with a sublinear regret upper bound.
    \item We provide two effective indicators $IND$ and $POD$ reflecting user behavior characteristics that can guide the search for appropriate embedding sizes from a statistical perspective.
    \item Experiments on four real-world datasets show that \textbf{DESS} outperforms the state-of-the-art methods in dynamic embedding size search for streaming recommender systems.
\end{itemize}

\section{Related Work}
\label{related}
%\subsection{Streaming Recommender System}
\textbf{Streaming Recommender System.} SRS is a kind of newly developed recommender system in coping with the high -throughput user data and their incremental properties~\citep{das2007google, song2008real, chen2013terec, devooght2015dynamic, wang2018streaming, chang2017streaming, he2023dynamically}.%~\citep{das2007google, song2008real, chen2013terec, song2017multi, devooght2015dynamic, wang2018streaming, chang2017streaming, he2023dynamically}. 
Different from the conventional recommender systems~\citep{zhang2016collaborative, he2017neural, chen2023sim2rec}, SRS needs to update its recommendation strategy in a dynamic manner to catch the user interest temporal dynamics. Early works~\citep{chandramouli2011streamrec, subbian2016recommendations},  known as memory-based methods, leverage similarity relationships in aggregated historical data to predict ratings. 
Some more advanced works%~\citep{rendle2008online, diaz2012real, devooght2015dynamic, chang2017streaming} 
~\citep{diaz2012real, devooght2015dynamic, chang2017streaming}propose to extend popular static recommender models like collaborative filtering and matrix factorization to the streaming fashion. However, most previous methods suffer from a common drawback: dividing the entire data stream into two segments, training on the former segment until convergence, and then testing on the latter segment, which is far from the real SRS scenario. In this work, we split the data of each time slice to two parts, train on the first part and test on the second part alternatively along the timeline, in such way to better fit the real streaming recommendation task.

%\subsection{Embedding Size Search}
\textbf{Embedding Size Search.} The embedding size search problem gradually attracts much attention because of the large models' increasing memory cost~\citep{covington2016deep, acun2021understanding} and the rapid development of neural architecture search techniques~\citep{liu2018darts, xie2018snas, brock2017smash, pham2018efficient, luo2018neural}. Some initial works focus on embedding size search on static recommender systems~\citep{joglekar2020neural, liu2021learnable, kang2020learning, ginart2021mixed, lyu2022optembed}. These approaches break the long-standing uniform-size embedding structure design. However, once determined, these embedding sizes can no longer be dynamically adjusted. As streaming recommender systems are more and more adopted, some recent works~\citep{zhaok2021autoemb, liu2020automated, zhao2021autodim, veloso2021hyper,zhao2021autodim} start to investigate the corresponding dynamic embedding size problem which means the embedding size for each ID is no more fixed. Generally speaking, previous methods can be divided into two mainstream schemes: \textit{soft-selection} and \textit{hard-selection}. In the \textit{soft-selection} scheme~\citep{zhaok2021autoemb, zhao2021autodim, liu2018darts}, the input of following representation learning layers is actually the weighted summation of transformed vectors corresponding to each embedding size. However, this category of methods suffer from the cold-start problem and the excessive memory cost. In the \textit{hard-selection} scheme~\citep{liu2020automated, veloso2021hyper}, only one size of embedding is selected at each time step. Nevertheless, previous works are still not reasonable enough for modeling the SRS update process. In addition, the algorithm performance lacks theoretical guarantee and needs to be improved if applied to the practical application. In this work, we follow the \textit{hard-selection} scheme and provide a more efficient dynamic embedding size search method.

%\subsection{Base Recommendation Model}
\textbf{Base Recommendation Model.} To make effective recommendations, a number of models have been proposed~\citep{ma2020probabilistic}, including matrix factorization (MF)-based models, distance-based models, and multi-layer perceptron-based models. Matrix factorization~\citep{koren2009matrix} is a representative and widely-used recommendation method which applies an inner product between the user and item embeddings to capture the interactions between users and items. Distance-based models, generally, compute the Euclidean distance between users and items for capturing fine-grained user preference~\citep{hsieh2017collaborative}. On the other hand, Neural Collaborative Filtering (NCF)~\citep{he2017neural}, a type of multi-layer perceptron-based method, models user-item interactions through neural network architectures, so that high-level nonlinearities within the user-item interaction can be learned. Following the settings in previous works~\citep{joglekar2020neural, zhaok2021autoemb, liu2020automated, zhao2021autodim}, we also choose NCF as the base recommendation model. Furthermore, we modify its architecture to fit the dynamic embedding size setting, which is detailed in Section~\ref{section:esann}. MF-based models and distance-based models are also explored to prove the wide applicability of our method.
%Following the settings in previous works~\citep{joglekar2020neural, zhaok2021autoemb, liu2020automated, zhao2021autodim}, we also choose NCF as the base recommendation model optimized by mean square error (MSE) loss and cross-entropy (CE) loss for the binary classification and multiclass classification task, respectively. The details of the model are introduced in Section \ref{section:esann}. Two other kinds of base models are also explored in Section \ref{section:furana}. 

\section{Preliminaries}

In this section, we first formalize the streaming recommendation problem. Then we derive the dynamic embedding size search task from the streaming model update process.

%\begin{figure}[ht]
%    \centering
%    \includegraphics[width=0.47\textwidth]{Figures/Fig1.png}
%    \caption{The right side shows the streaming recommender system (SRS) workflow. The left side illustrates how the streaming %model update module \textbf{SMU} works.
%    }
%    \label{fig:SRS}
%\end{figure}

\begin{figure*}
  \includegraphics[width=0.92\textwidth]{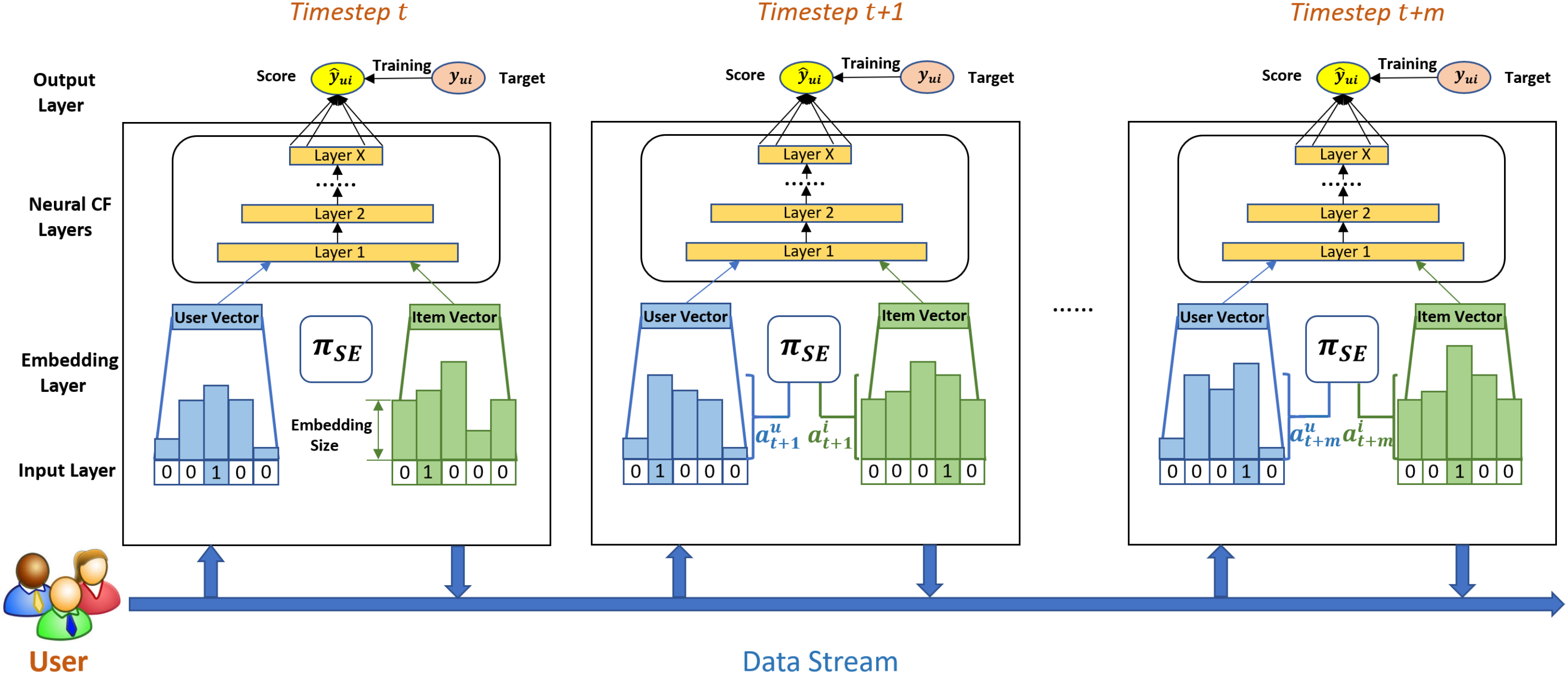}
  \caption{Illustration of the dynamic embedding size search process. At each timestep, the embedding layer outputs embedding vectors with sizes determined by the $\pi_{SE}$. After transformation, they will be input into Neural CF layers for further inference.}
  \label{fig:DESS}
  \vspace{-0.25cm}
\end{figure*}

%$\mathcal{K}:$ Number of arms which are embedding size candidates in this setting.\\
%$\mathcal{T}:$ Total number of timesteps, $\mathcal{t}$ is the corresponding timestep index ranging from 1 to %$\mathcal{T}$.\\
%$N:$ The size of mini-batch $D_{\mathcal{t}}$. $k$ is the corresponding index ranging from 1 to N.\\
%$T:$ The length of the data stream, equal to the product of $N$ and the aforementioned $\mathcal{T}$. $t$ is the corresponding data point index ranging from 1 to $T$. In the following algorithm parts, we will use variables with $va$ as superscript and $t$ as subscript like $x^{va}_t$ and $a^{va}_t$ which means $t_{th}$ data in the validation data stream. Note that the validation data stream is not entirely consistent with the real data stream. \\
%$\mathcal{C}:$ The set of contexts. In this setting, the contexts are feature groups that determine the optimal embedding sizes for the corresponding users or items.\\
%$\pi^{u}_{SE}$: The dynamic embedding size search policy which selects the suitable user embedding sizes dynamically in the data flow.\\
%$\pi^{i}_{SE}$: The dynamic embedding size search policy which selects the suitable item embedding sizes dynamically in the data flow.\\
%$r_t:$ The reward received by the policy at data point $t$.\\
%$\theta_{t, a}:$ The updated parameter of reward model for arm $a$ at data point $t$.\\
%$\theta^*_{t, a}:$ The parameter of oracle reward model  $\mathcal{R}$ for arm $a$ at data point t.\\
%$M_{\mathcal{t}}:$ The updated recommender model at timestep $\mathcal{t}$.\\

\subsection{Streaming Recommendation}
\label{section:StreamRec}

One prominent advantage of SRS is that they can update and respond instantaneously for catching users' intentions and demands~\citep{chang2017streaming, chandramouli2011streamrec}. Due to the high volume of online data, previous works~\citep{he2023dynamically, wang2020streaming, wang2022streaming, zhaok2021autoemb, liu2020automated} split the user-item interaction stream into short-term segments. Following this setting, we split the whole data stream of length $L$ into $T$ consecutive segments $D_1,...,D_t,...,D_T$ with the same length $|D_t|$ ($L = |D_t| \times T$). Each segment $D_t$ is then divided into the training part $D^{tr}_t$ and test part $D^{te}_t$. On this basis, the streaming recommendation task is formulated as: given $D^{tr}_{1}, D^{tr}_{2},..., D^{tr}_{t},..., D^{tr}_{T}$, it is supposed to train a model $M$ to predict the user preference in $D^{te}_{1}, D^{te}_{2},..., D^{te}_{t},..., D^{te}_{T}$, where $D^{tr}_{t} \cup D^{te}_{t} = D_{t}$ and $D^{tr}_{t} \cap D^{te}_{t} = \emptyset$. The overall performance is evaluated by the average recommendation accuracy over the entire timeline. 

%\begin{figure}[ht!]
%    \centering
%    \includegraphics[width=0.4\textwidth]{Figures/Fig3.png}
%    \caption{The illustration of data distribution shift across different timesteps
%    }
%    \label{fig:DDS}
%    \vspace{-0.3cm}
%\end{figure}

\subsection{Dynamic Embedding Size Search}

With the wide use of deep recommendation models, embeddings are largely investigated to represent users, items, and other auxiliary features. However, the conventional design of setting identical and static embedding sizes suffers from the unsatisfying model prediction performance and the unacceptable memory cost. To solve these problems, many works~\citep{joglekar2020neural, hutter2011sequential,liu2020automated, zhao2021autodim, veloso2021hyper} are proposed for the embedding size search. To make the search fit into the streaming scenario, a \textbf{streaming update process} should be first introduced: the model $M$ inherits the parameters from the previous moment $M_{t-1}$ and updates itself to $M_t$ with the current training data $D^{tr}_{t}$. Following these, %and the streaming recommendation task description in Section~\ref{section:StreamRec}, 
the dynamic embedding size search task can be further formulated: optimizing embedding size search policies $\pi^u_{SE}$ and $\pi^i_{SE}$ which accordingly control user and item embedding sizes of recommendation model $M$ at each timestep, so that the overall model performance can be satisfying (see Figure~\ref{fig:DESS}). For the ease of illustration, $\pi_{SE}$ refers to both $\pi^{u}_{SE}$ and $\pi^{i}_{SE}$.

%The major notations are summarized in Appendix~\ref{section:notations}.

\section{Methodology}
In this section, we introduce our approach for dynamic embedding size search in streaming recommender systems. First, we model the embedding size search as a bandit problem and formalize the objective. Then we analyze and quantify the characteristics that can determine optimal embedding sizes from a statistical perspective. Next, we elaborate the non-stationary LinUCB-based \textbf{DESS} method (shown in Algorithm~\ref{alg:DESS}) to conduct the dynamic embedding size search, and provide the corresponding theoretical guarantee analysis. Finally, we introduce the structure of our streaming recommendation model---an embedding size adaptive neural network.

\subsection{Embedding Size Search as Bandits}
%\subsection{Problem Setting and Formulation}
\label{section:promod}
% Bandit problem is a kind of online sequential decision making problem in which a fixed limited set of resources must be allocated between competing choices in a way that maximizes  the sum of rewards earned through a sequence of lever pulls~\citep{lattimore2020bandit}. Similarly, the target of the embedding size searcher is to optimize the average/cumulative model performance by selecting different embedding sizes from the candidate list at different timesteps.  Therefore, we model the dynamic embedding size search as a bandit problem and the objective of the searcher is to minimize the expected dynamic \textbf{pseudo-regret} defined as:

The target of the embedding size search in streaming scenarios is to optimize the average/cumulative model performance by selecting appropriate embedding sizes at different timesteps. From the temporal view, the search process is, in nature, a sequence of size value decisions according to data characteristics. To solve this sequential decision-making problem, Multi-armed Bandits (MAB) are a promising approach where a fixed limited set of resources must be allocated between competing choices in a way that maximizes the sum of rewards earned through a sequence of lever pulls~\citep{lattimore2020bandit}. Therefore, to consider the model's recommendation performance at each timestep, we model the dynamic embedding size search as a bandit problem, and the objective of the search policy $\pi_{SE}$ is to minimize the expected dynamic \textbf{pseudo-regret} defined as:
\begin{equation}
\begin{aligned}
\Bar{R}_{T} = \underset{\pi: C\to{1,...,K}}{\mathop{max}}
\mathbb{E}[\underset{t=1}{\sum^{T}} \mathcal{L}^{te}_{t}(\pi_{SE}) - \underset{t=1}{\sum^{T}}\mathcal{L}^{te}_{t}(\pi)],
\end{aligned}  
\label{pseudo-regret}
\end{equation}
where $\mathcal{L}^{te}_{t}(\pi_{SE})$ is the batch loss of $M$ with embedding sizes determined by search policy $\pi_{SE}$ on test data $D^{te}_{t}$. $\mathcal{L}^{te}_{t}(\pi)$ is the model test loss received from pulling the arm that an arbitrary policy $\pi$ recommends at the current state.  $C$ is the set of context information that can help the policy $\pi$ select the best arm from the arm candidates $1,...,K$ at different timesteps. Note that the ideal $\Bar{R}_{T}$ is obtained when $\pi$ is the optimal one. Thus, the pseudo regret for $\pi_{SE}$ is the difference between the actual loss it incurs and the loss incurred by the best possible embedding size search policy~\citep{lattimore2020bandit}.

Since the test data is actually inaccessible in the training phase, we utilize the validation data to interact with the bandit directly and update the search policy. Especially, in the streaming recommendation scenario, the union of training data and test data at the last timestep $t-1$ ($D_{t-1}$) can be regarded as the validation data for timestep $t$~\citep{zhaok2021autoemb, liu2020automated}. Then, the corresponding dynamic $\textbf{regret}$ is:
\begin{equation}
\begin{aligned}
R_{T} = \underset{\pi:C\to{1,...,K}}{\mathop{max}} \underset{t=1}{\sum^T}\underset{j=1}{\sum^{|D_{t-1}|}} r^{va}_{t,j}(\pi(\mathbf{x})) - \underset{t=1}{\sum^T}\underset{j=1}{\sum^{|D_{t-1}|}}
 r^{va}_{t,j}(\pi_{SE}(\mathbf{x})),
\end{aligned}    
\label{regret}
\end{equation}
where $r^{va}_{t,j}(\pi(\mathbf{x}))$ is the reward received from pulling the arm that the $\pi$ recommends in the validation phase. $r^{va}_{t,j}(\pi_{SE}(\mathbf{x}))$ is the reward actually received by our $\pi_{SE}$ on validation data. The $\mathbf{x}$ indicates the context as the input of the bandit model, whose details will be provided in Section~\ref{section:infana}. Here, we further explain the reward $r_{t,j}$, noted as $r_l (l= (t-1) \times |D_t| + j, 1 \leq l \leq L)$  and the arm $a$.

\textbf{Reward}. The reward $r_l$ is defined based on the performance $\mathcal{L}^{new}_{l}$ of the temporarily updated embedding structure by $\pi_{SE}$ and the previous structure's performance $\mathcal{L}^{old}_{l}$ on $l$-th interaction of the validation data stream. To fairly compare the effectiveness of such two embedding structures with different sizes, we temporarily tune their parameters with the $l$-th interaction. $r_l$ here is designed as a binary variable and can only be $0$ or $1$. The formula is following:
\begin{equation}
r_l = \left \{
\begin{array}{rcl}
     1, &if& \mathcal{L}^{old}_{l} - \mathcal{L}^{new}_{l} > threshold\\
     0, &if& \mathcal{L}^{old}_{l} - \mathcal{L}^{new}_{l} < threshold
\end{array}  
\right. %\}
\label{equ:reward}
\end{equation}

In the real-world application, $r_l$ can be designed as a continuous real number $\mathcal{L}^{old}_{l} - \mathcal{L}^{new}_{l}$ for performance optimization. 

\textbf{Arm}. The arms $a$ here are actually different embedding size candidates or other embedding size adjustment operations, like increasing or decreasing embedding sizes.

\subsection{Embedding Size Indicator}
\label{section:infana}
% To effectively determine the size of embeddings, the indicator to increase or decrease embedding size is inevitable to explore. We make an analysis to which features in the user-item interaction determine the optimal embedding size for each user or item at data point $t$. According to the statistics~\citep{dunteman1989principal, jambulapati2020robust, battaglino2020generalization}, the greater the data dispersion degree, the greater the information amount it contains which means we need a longer embedding vector to describe the historical information. Based on this, we quantify the information amount of historical data. Assume we have a set of $D$-dimensional raw feature vectors $\textbf{f}_1, \textbf{f}_2,..,\textbf{f}_M$ corresponding to each item. Till data point index $l$, let user $u$ have rated a subset of the items indexed by $i_1, i_2,...,i_P$, then the interest diversity of user $u$ can be formulated with centroid diameter distance:
To effectively determine the sizes of embeddings, the indicator to increase or decrease embedding sizes is worth exploring because only browsing frequency is not sufficient to prompt the search policy to make the correct decision. According to~\citep{dunteman1989principal, jambulapati2020robust, battaglino2020generalization}, a larger data dispersion degree indicates a greater amount of information the data contains. Thus, when the data dispersion degree is large, the size of the embedding should be large to represent the whole historical information. Motivated by this, we follow a similar fashion to quantify the amount of information in historical data by leveraging the explicit item features independent of the user-item interactions. Assume we have a set of raw feature vectors $\mathbf{F}_1, \mathbf{F}_2,..,\mathbf{F}_N$ corresponding to each item. Till $l$-th interaction of the data stream, let user $u$ have rated a subset of the items indexed by $i_1, i_2,...,i_H$, then the interest diversity of user $u$ can be formulated by the centroid diameter distance:
\begin{equation}
\begin{aligned}
IND^u_l = \frac{1}{H} \underset{h=1}{\sum^H}  \sqrt{(\mathbf{F}_{i_h} - \mathbf{Q}^u_l)(\mathbf{F}_{i_h} - \mathbf{Q}^u_l)^\top},
\end{aligned}    
\label{equ:ind}
\end{equation}
where $\mathbf{Q}^u_l$ is the mean vector of $\mathbf{F}_{i_1}, \mathbf{F}_{i_2},...,\mathbf{F}_{i_H}$ and represents the user $u$'s mean interest. For item $i$, assume it has been rated by users $u_1, u_2,.., u_H$ till $l$-th interaction, the diversity of its property is:
\begin{equation}
\begin{aligned}
POD^i_l = \frac{1}{H}\underset{h=1}{\sum^H}  \sqrt{(\mathbf{Q}^{u_h}_l - \mathbf{P}^i_l)(\mathbf{Q}^{u_h}_l - \mathbf{P}^i_l)^\top},
\end{aligned}    
\label{equ:pod}
\end{equation}
where $ \mathbf{P}^i_l$ is the mean vector of $\mathbf{Q}^{u_1}_l, \mathbf{Q}^{u_2}_ l,...,\mathbf{Q}^{u_H}_l$ and represents item $i$'s mean property.  In this way, we define the context $\mathbf{x}_l$ for the user embedding size search policy $\pi^{u}_{SE}$ as the combination of frequency and information diversity $(FRE^u_l, IND^u_l)$, where $FRE^u_l$ is the occurrence number of user $u$ in historical data. The context for the item embedding size search policy $\pi^{i}_{SE}$ is defined as $(FRE^i_ l, POD^i_l)$ similarly, where $FRE^i_l$ is the occurrence of item $i$ in historical data.

\subsection{Non-stationary LinUCB-based Search Policy}
\label{section:bansea}
Despite the effectiveness of linear MAB%~\citep{abbasi2011improved, auer2002using}
, some recent works~\citep{russac2019weighted, wu2018learning, kim2020randomized, zhao2020simple} focus on a more general setting: the constraint that requires fixed optimal regression parameters $\bm{\theta}^{*}$ is relaxed, which is more suitable for our scenario. To better balance the exploration and exploitation in such a setting, we design our non-stationary LinUCB-based \textbf{DESS} algorithm for dynamic embedding size search. Its effectiveness on memory cost and time efficiency will be elaborated in Section~\ref{sec:exp}. Due to the fact that the accumulated historical data for each user and item can only become richer and richer as data streams in, we simplify the embedding size search as a binary selection problem, where $\pi_{SE}$ only needs to decide if increasing the embedding size to the subsequent larger size candidate. 

The algorithm details are described in Algorithm \ref{alg:DESS}. Generally, the whole algorithm is separated to two parts: \textit{Updating Non-stationary LinUCB-based Search Policy $\pi_{SE}$} and \textit{Updating recommendation model}, which are executed one after the other at each timestep $t$. Different arms $a$ in our method share the common context information $x_l$ about the frequency and interest/property diversity: $\mathbf{x}_{l,1} = \mathbf{x}_{l,2} = ... = \mathbf{x}_{l, K} = \mathbf{x}_l$. Based on the assumption that the expected payoff $r_l$ is linear to its context $\mathbf{x}_l \in \mathbb{R}^d $ , we set disjoint linear reward models $\bm{\theta}_1, \bm{\theta}_2, ..., \bm{\theta}_{K}$ for corresponding arms $1,2...,K$ to estimate rewards when selecting different arms. Note that in this paper, $\bm{\theta}_{l,a}$ indicates the parameter of $\bm{\theta}_{a}$ at the $l$-th user-item interaction. For such disjoint linear models, we use ridge regression~\citep{marquardt1975ridge} to solve them. And the objective is to minimize the regularized weighted residual sum of squares, thus the $\hat{\bm{\theta}}_{l,a}$ is defined as follows:
\begin{equation}
\begin{aligned}
\underset{\bm{\theta} \in \mathbb{R}^d}{arg min} (\underset{s=1}{\sum^l}
 \mathbbm{1}(a_s=a)\gamma^{l-s}(r_s - \langle \mathbf{x}_{s,a}, \bm{\theta}\rangle)^2 + \lambda \Vert \bm{\theta} \Vert^2_2 ),
 \label{equ:objective}
\end{aligned}    
\end{equation}
where $\langle,\rangle$ indicates the inner product operation, and $a_s$ is the arm selected by the $\pi_{SE}$ at $s$-th interaction ($1\leq s \leq l$). Eq.~\ref{equ:objective} is actually the regularized weighted least-squared estimator of $\bm{\theta}_a^*$ at $l$-th interaction. The conduct of the weighted forgetting mechanism (discount factor $\gamma$) is to reduce the interference from the outdated data~\citep{besbes2014stochastic, russac2019weighted, wu2018learning} and help $\pi_{SE}$ pay more attention to the recent user/item behaviors. Furthermore, we have the solution for Eq.~\ref{equ:objective}:
\begin{equation}
\begin{aligned}
\hat{\bm{\theta}}_{l,a} &= \mathbf{V}^{-1}_{l,a}\underset{s=1}{\sum^l}\mathbbm{1}(a_s=a)\gamma^{l-s}\mathbf{x}_{s,a} r_s,\\ where \quad \mathbf{V}_{l,a} &= \underset{s=1}{\sum^l}\mathbbm{1}(a_s=a)\gamma^{l-s}\mathbf{x}_{s,a} \mathbf{x}^{\top}_{s,a} + \lambda \mathbf{I}_d,
\end{aligned}    
\end{equation}
and $\mathbf{I}_d$ denotes the $d$-dimensional identity matrix. Here, similar to $\mathbf{V}^{-1}_{l,a}$, we define a matrix $\widetilde{\mathbf{V}}_{l,a}$ as an intermediate variable in our algorithm to help obtain the confidence ellipsoid:
\begin{equation}
\begin{aligned}
\widetilde{\mathbf{V}}_{l,a} &= \underset{s=1}{\sum^l}\mathbbm{1}(a_s=a)\gamma^{2(l-s)}\mathbf{x}_{s,a} \mathbf{x}^{\top}_{s,a} + \lambda \mathbf{I}_d,
\end{aligned}    
\end{equation}
which is strongly connected to the variance of the estimator $\hat{\bm{\theta}}_{l,a}$. Applying the online version of ridge regression~\citep{russac2019weighted, kim2020randomized, zhao2020simple}, the update formulations of $\mathbf{V}_{a_l}, \widetilde{\mathbf{V}}_{a_l}, \hat{\bm{\theta}}_{a_l}$ are shown as follows:
\begin{equation}
\begin{aligned}
\mathbf{V}_{a_l} &= \gamma \mathbf{V}_{a_l} + \mathbf{x}_{l, a_l} \mathbf{x}^\top_{l, a_l} + (1-\gamma)\lambda \mathbf{I}_d, \\ \widetilde{\mathbf{V}}_{a_l} &= \gamma^2 \widetilde{\mathbf{V}}_{a_l} + \mathbf{x}_{l, a_l} \mathbf{x}^\top_{l, a_l} + (1-\gamma^2)\lambda \mathbf{I}_d, \\ \mathbf{b}_{a_l} &= \gamma \mathbf{b}_{a_l} + r_l \mathbf{x}_{l, a_l}, \hat{\bm{\theta}}_{a_l} = \mathbf{V}^{-1}_{a_l} \mathbf{b}_{a_l},
\label{equ:update}
\end{aligned}    
\end{equation}
where $\mathbf{b}$ is an intermediate variable to help compute $\hat{\bm{\theta}}$. The initialization of $\mathbf{V}_{a}, \widetilde{\mathbf{V}}_{a}, \hat{\bm{\theta}}_{a}$ for each arm $a$ is provided in the \textbf{Initialize} part of Algorithm \ref{alg:DESS}. During the algorithm execution, we use Eq.~\ref{equ:update} to update such variables for the selected arm at each interaction.

Finally, we introduce how our non-stationary LinUCB-based policy $\pi_{SE}$ selects the most promising arm. Following related works~\citep{russac2019weighted, wu2018learning, kim2020randomized, zhao2020simple}, we first obtain the confidence value $\beta_l$ (coefficient of confidence ellipsoid) which controls the exploration level:
\begin{equation}
\begin{aligned}
\beta_{l} = \sqrt{\lambda}S + \sigma \sqrt{2\log(\frac{1}{\delta}) + d\log(1 + \frac{U^2(1-\gamma^{2l})}{\lambda d(1-\gamma^2)})},
\end{aligned}    
\end{equation}
where $S$ is the upper bound for parameters ($\forall l,a, \Vert \bm{\theta}_{l,a}^*\Vert_2 \leq S$), $U$ is the upper bound for contexts ($\forall l,a, \Vert \mathbf{x}_{l,a}\Vert_2 \leq U$), $\sigma$ is the subgaussian constant, and $\delta$ is a pre-designated probability. Based on this, we can derive the upper confidence bound (UCB) which considers both the estimated reward and the uncertainty (confidence ellipsoid) of the reward estimation, in such a way to better balance the exploration and exploitation for arm selection:
\begin{equation}
\begin{aligned}
UCB(a) = \mathbf{x}_{l, a}^{\top} \hat{\bm{\theta}}_{a} + \beta_l \sqrt{\mathbf{x}^{\top}_{l,a} \mathbf{V}_a^{-1} \widetilde{\mathbf{V}}_a \mathbf{V}_a^{-1} \mathbf{x}_{l,a}}.
\end{aligned}    
\end{equation}
After computing $UCB(a)$ for each arm, $\pi_{SE}$  will select the arm with the highest UCB score as the embedding size control command.

\begin{algorithm}[t]
 \caption{Dynamic Embedding Size Search (DESS)}
 \label{alg:DESS}
 \begin{flushleft}
  \textbf{Input}:\\
  $\eta$ (learning rate for recommender model), probability $\delta$, subgaussianity constant $\sigma$, context dimension $d$, regularization $\lambda$,
  upper bound for contexts $U$, upper bound for parameters $S$, discount factor $\gamma$\\
 \textbf{Initialize:} initial recommender model $M_0$, $\mathbf{b}_a = \mathbf{0}_{\mathbb{R}^d}, \mathbf{V}_a = \lambda \mathbf{I}_d, \widetilde{\mathbf{V}_a} = \lambda \mathbf{I}_d, \hat{\bm{\theta}}_a = \mathbf{0}_{\mathbb{R}^d}$ for each arm $a$,\\
 user-item interaction data stream $\{ (D^{tr}_1, D^{te}_1),..., (D^{tr}_{T}, D^{te}_{T})\}$ which contains $T$ segments of data in chronological order\\
 \textbf{Process}:
 \end{flushleft} 
 \begin{algorithmic}[1]
 \For{each $t$ = 1,2,...,$T$}
 \State $/*$ Update Non-stationary LinUCB-based Policy $\pi_{SE}$ $*/$
 \State Collect the last segment of data $(D^{tr}_{t-1}, D^{te}_{t-1})$
  \For{each interaction in $(D^{tr}_{t-1} \cup D^{te}_{t-1})$}
    %\State Receive context $\mathbf{x}^{val}_{l,a}$ for each arm
    %\State $\beta_{l} = \sqrt{\lambda}S + \sigma \sqrt{2\log(\frac{1}{\delta}) + d\log(1 + \frac{U^2(1-\gamma^{2l})}{\lambda d(1-\gamma^2)})}$
    %\State $UCB(a) = \mathbf{x}_{l, a}^{{val}^\top} \hat{\bm{\theta}}_{a} + \beta_l \sqrt{\mathbf{x}^{{val}^\top}_{l,a} \mathbf{V}_a^{-1} \widetilde{\mathbf{V}}_a \mathbf{V}_a^{-1} \mathbf{x}^{val}_{l,a}}$ for each $a \in \mathcal{A}^{val}_l$
    %\State $a^{val}_l = argmax_{a \in \mathcal{A}^{val}_l}(UCB(a))$
    %\State Temporarily change embedding sizes according to $a^{val}_l$
    %\State Input interaction into $M_{t-1}$ and receive reward $r_l$
    %\State \textbf{Updating phase:} $\mathbf{V}_{a^{val}_l} = \gamma \mathbf{V}_{a^{val}_l} + \mathbf{x}^{val}_{l, a^{val}_l} \mathbf{x}^{{val}^\top}_{l, a^{val}_l} + (1-\gamma)\lambda \mathbf{I}_d,$ $\widetilde{\mathbf{V}}_{a^{val}_l} = \gamma^2 \widetilde{\mathbf{V}}_{a^{val}_l} + \mathbf{x}^{val}_{l, a^{val}_l} \mathbf{x}^{{val}^\top}_{l, a^{val}_l}  + (1-\gamma^2)\lambda \mathbf{I}_d, \mathbf{b}_{a^{val}_l} = \gamma \mathbf{b}_{a^{val}_l} + r_l \mathbf{x}^{val}_{l, a^{val}_l}, \hat{\bm{\theta}}_{a^{val}_l} = \mathbf{V}^{-1}_{a^{val}_l} \mathbf{b}_{a^{val}_l}$
    \State Receive context $\mathbf{x}_{l,a}$ for each arm $a$
    \State $\beta_{l} = \sqrt{\lambda}S + \sigma \sqrt{2\log(\frac{1}{\delta}) + d\log(1 + \frac{U^2(1-\gamma^{2l})}{\lambda d(1-\gamma^2)})}$
    \State $UCB(a) = \mathbf{x}_{l, a}^{\top} \hat{\bm{\theta}}_{a} + \beta_l \sqrt{\mathbf{x}^{\top}_{l,a} \mathbf{V}_a^{-1} \widetilde{\mathbf{V}}_a \mathbf{V}_a^{-1} \mathbf{x}_{l,a}}$ for each $a$
    \State $a_l = argmax_{a \in \mathcal{A}}(UCB(a))$
    \State Temporarily change embedding sizes according to $a_l$
    \State Temporarily tune embedding parameters 
    \State Input interaction into $M_{t-1}$ and receive reward $r_l$
    \State \textbf{Updating:} $\mathbf{V}_{a_l} = \gamma \mathbf{V}_{a_l} + \mathbf{x}_{l, a_l} \mathbf{x}^\top_{l, a_l} + (1-\gamma)\lambda \mathbf{I}_d$,  $\widetilde{\mathbf{V}}_{a_l} = \gamma^2 \widetilde{\mathbf{V}}_{a_l} + \mathbf{x}_{l, a_l} \mathbf{x}^\top_{l, a_l} + (1-\gamma^2)\lambda \mathbf{I}_d, \mathbf{b}_{a_l} = \gamma \mathbf{b}_{a_l} + r_l \mathbf{x}_{l, a_l}, \hat{\bm{\theta}}_{a_l} = \mathbf{V}^{-1}_{a_l} \mathbf{b}_{a_l}$
  \EndFor
  
  \State
 \State $/*$ Update Recommendation Model $M$ $*/$
 \State Collect the current segment of data $(D^{tr}_{t}, D^{te}_{t})$
  \State Output actions $a_{t}$ from $\pi_{SE}$
  \State Permanently change the embedding sizes for user-item pairs in $D^{tr}_{t}$ according to  $a_{t}$ 
  \State Input $D^{tr}_{t}$ to $M_{t-1}$ and update the model to $M_{t}$ 
  \State Report the accuracy and test loss of $M_{t}$ on test data $D^{te}_{t}$
 
 \EndFor
 \end{algorithmic}
\end{algorithm}

\subsection{Theoretical Analysis}
\label{section:theory}
We provide the upper regret bound analysis for \textbf{DESS} in this section. As far as we know, this is the first theoretical regret bound for the non-stationary contextual linear bandit with disjoint arm-associated parameter vectors $\bm{\theta}_a$. 
\begin{definition}{\textbf{Parameter Variation Budget.}}
For the search policy, the true oracle parameters $\{\bm{\theta}^{*}_{l,a}\}^{L}_{l=1}$ are actually unknown. And their shifts can be quantified by the \textit{variation budget} which measures the magnitude of non-stationarity in the dynamical data stream. This can be defined as:
\begin{equation}
    B^{*}_L := \underset{l=1}{\sum^{L-1}}  \underset{a}{max} \Vert \bm{\theta}^{*}_{l+1,a} - \bm{\theta}^{*}_{l,a}\Vert_2.
\end{equation}
\end{definition}
\begin{assumption}{\textbf{Variation Budget Upper Bound}}
\\
We assume that the variation budget is bounded by a known quantity $B_L$ similar to the previous literature~\citep{besbes2014stochastic, russac2019weighted, kim2020randomized, zhao2020simple}, that is $B^{*}_L \leq B_L$.
\end{assumption}
\begin{assumption}{\textbf{Bounded Reward}}
\\
We assume that the reward $r_l$ is bounded by the subgaussianity constant $2\sigma$, $0 \leq r_l \leq 2\sigma$ which can be easily satisfied. For example, in the above algorithm description part, the $\sigma$ can be set as $0.5$.
\end{assumption}

\begin{lemma}
Let prediction error $Er(\mathbf{x}_{l,a}, \bm{\theta}_{l,a}) = |\langle \mathbf{x}_{l,a}, \bm{\theta}^*_{l,a}\rangle - \langle \mathbf{x}_{l,a}, \bm{\theta}_{l,a}\rangle|,\\
k=\underset{\mathbf{x}, \bm{\theta}}{sup} \langle \mathbf{x}, \bm{\theta} \rangle, c=\underset{\mathbf{x}, \bm{\theta}}{inf} \langle \mathbf{x}, \bm{\theta} \rangle, $ and $D \in \mathbb{N}^*$. With probability at least $1 - \delta$: for all $a \geq 1$, the following holds:\\
%\begin{equation}
   $Er(\mathbf{x}_{l,a}, \bm{\theta}_{l,a}) \leq \frac{2k}{c}\beta_l \Vert \mathbf{x}_{l,a} \Vert_{\mathbf{V}^{-1}_{l,a}} + \frac{2kU}{c} \sqrt{1 + \frac{L^2}{\lambda(1-\gamma)}}(\frac{2kS U^2}{\lambda} \frac{\gamma^D}{1-\gamma} + \\
     k\sqrt{\frac{d}{\lambda(1-\gamma)}} \sum_{s=l-D}^{l-1} \Vert \bm{\theta}^*_{s,a}-\bm{\theta}^*_{s+1,a}\Vert_2)$.
%\end{equation}
\label{lemma1}
\end{lemma}
%\\
\begin{lemma}
%Similar to~\citep{russac2019weighted}, let $\{\mathbf{x}_{s,a^{*}_s}\}^{L}_{s=1}$ a sequence in $\mathbb{R}^d$ such that $\Vert \mathbf{x}_{s,a^{*}_s} \Vert_2 \leq U$ for all $s \in \mathbb{N}^{*}$, and let $\lambda$ be a non-negative scalar. For $l \geq 1$, define $V_{l, a^{*}_l} := \sum_{s=1}^{l} \gamma^{l-s} \mathbf{x}_{s, a^{*}_s} \mathbf{x}_{s, a^{*}_s}^T + \lambda \mathbf{I}_d$. $a^{*}_l$ is the optimal arm that $\pi_{SE}$ should select. The following inequality holds:\\
Similar to~\citep{russac2019weighted}, let $\{\mathbf{x}_{s,a^{*}_l}\}^{l}_{s=1}$ a sequence in $\mathbb{R}^d$ such that $\Vert \mathbf{x}_{s,a^{*}_l} \Vert_2 \leq U$ for all $s \in \mathbb{N}^{*}$. $a^{*}_l$ is the optimal arm that $\pi_{SE}$ should select at $l$-th interaction. For $l \geq 1$, define $V_{l, a^{*}_l} := \sum_{s=1}^{l} \mathbbm{1}(a_s=a^{*}_l) \gamma^{l-s} \mathbf{x}_{s, a^{*}_l} \mathbf{x}_{s, a^{*}_l}^{\top} + \lambda \mathbf{I}_d$. Given $\lambda \geq 0$, the following inequality holds:\\
%\begin{equation}
   $ \sum_{l=1}^L \Vert \mathbf{x}_{l,a^*_l} \Vert_{\mathbf{V}^{-1}_{l,a^*_l}} \leq 2\max(1, L^2/\lambda)(dL\log(\frac{1}{\gamma}) + d\log(1+\frac{U^2(1-\gamma^L)}{\lambda d(1-\gamma)}))$.
%\end{equation}
\label{lemma2}
\end{lemma}
%\\
\vspace{-2mm}
\begin{theorem}
Under the assumptions above, the regret of the \textbf{DESS} algorithm is bounded for all $\gamma \in (0,1)$ and integer $D \geq 1$, with the probability at least $1-\delta$, by \\
    $R_L \leq \sqrt{32\max(1,U^2/\lambda)}\frac{k}{c}\beta_L\sqrt{dL}\sqrt{L\log(\frac{1}{\gamma})+\log(1+\frac{U^2(1-\gamma^L)}{\lambda d(1-\gamma)})} + \\
    \frac{8 k^2 S U^3 \gamma^D}{c\lambda(1-\gamma)}L + \frac{8 k^2 S U^4 \gamma^D}{c \lambda^{\frac{3}{2}} (1-\gamma)^{\frac{3}{2}}}L + \frac{4 k^2 UD}{c\sqrt{\lambda}}\sqrt{\frac{d}{1-\gamma}}B_L + \frac{4 k^2 U^2 D}{c\lambda} \frac{\sqrt{d}}{1-\gamma}B_L$.
\label{theo1}
\end{theorem}

\begin{corollary}
By choosing discount factor $\gamma = 1 - (B_L/(\sqrt{d}L))^{2/5}$, the regret of \textbf{DESS} algorithm is asymptotically upper bounded with high probability by a term $\mathcal{O}(d^{9/10}B_L^{1/5}L^{4/5})$ when $L \rightarrow \infty$.
\label{corollary1}
\end{corollary}
The detailed proofs of theorem \ref{theo1} and corollary \ref{corollary1} are provided in Appendix~\ref{section: proof of theorem 1} and ~\ref{section: proof of corollary 1}, respectively.

\subsection{Embedding Size Adaptive Neural Network}
\label{section:esann}
\begin{figure*}
    \centering
    \includegraphics[width=0.85\textwidth]{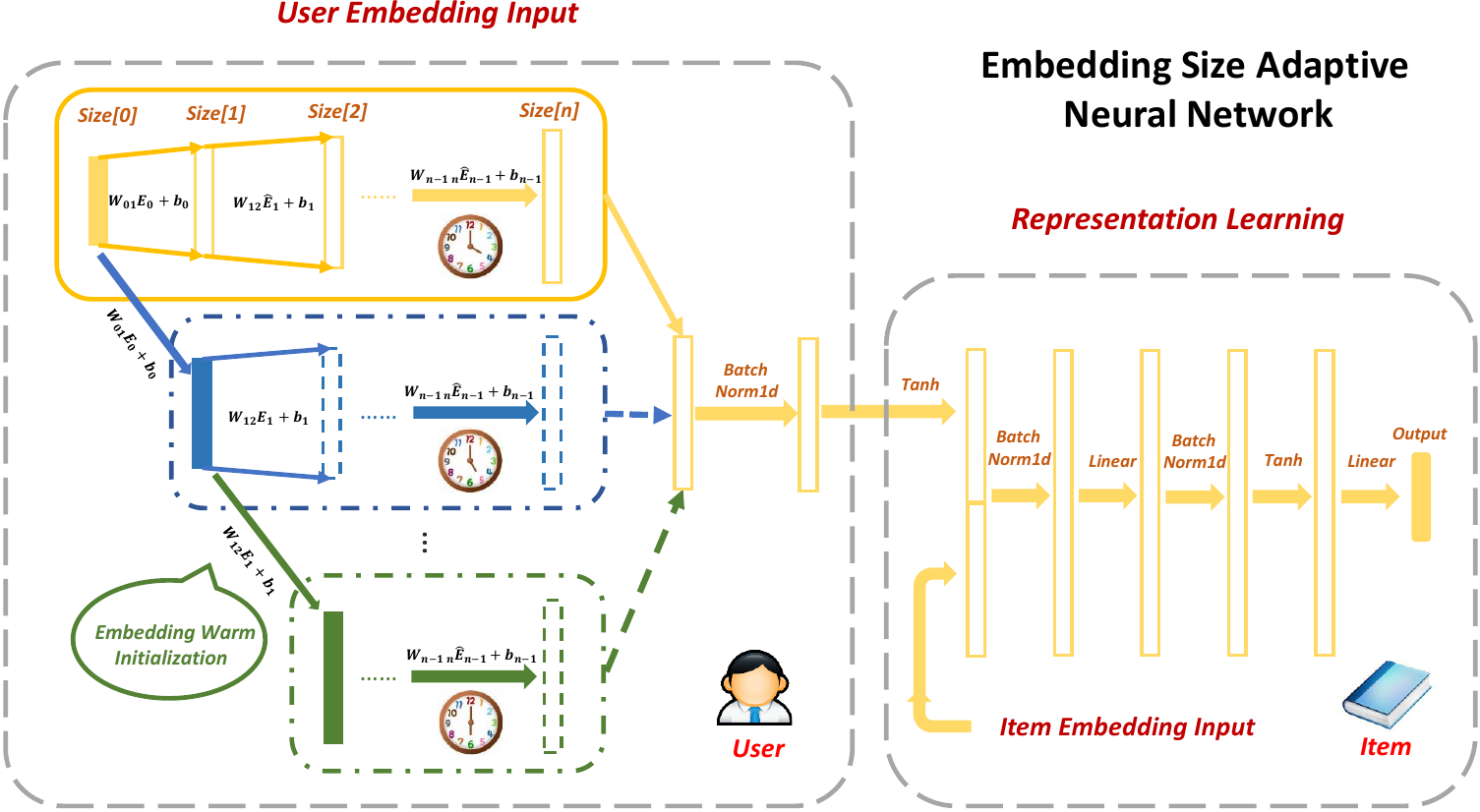}
    \caption{The illustration of \textit{embedding size adaptive neural network} structure. Only the user \textit{embedding input} is shown in detail above. The item \textit{embedding input} structure is similar. Different colors indicate the model inference process at different timesteps. The dashed rectangles represent the tensors that do not participate in the current forward propagation. The hollow rectangles represent intermediate tensors that will not be saved after the neural network forward propagation.
    }
    \label{fig:DNN}
    \vspace{-0.3cm}
\end{figure*}
\subsubsection{Model Inference}
\label{subsection:mi}
As mentioned in Section~\ref{related}, based on the NCF model~\citep{he2017neural,joglekar2020neural, liu2020automated}, we design an \textit{embedding size adaptive neural network} shown in Figure \ref{fig:DNN} as the streaming recommendation model $M$ in Algorithm \ref{alg:DESS}. Different from the conventional design that assigns one or a set of embedding sizes for each user or item in advance~\citep{he2017neural, wang2019neural, zhang2016collaborative, liu2018darts, liu2020automated}, the embedding size of each user or item can be selected flexibly from a group of size candidates at each timestep in our proposed structure. Suppose that the embedding size group for users and items are both $ size = [ s_0, s_1, .., s_{n}] (s_0 < s_1...< s_{n})$  for simplicity. Actually, the candidates for each user or item can be different. The initial size for each ID is set as the minimum value $s_0$ of the candidate group. $\{\mathbf{W}_{01}, \mathbf{b}_0, \mathbf{W}_{12}, \mathbf{b}_1, ..., \mathbf{W}_{n-1 n}, \mathbf{b}_n\}$ is a sequence of linear transformation parameters that will unify the embedding size to $s_n$ before inputting it to the \textit{representation learning} part. Assume that the current embedding size for user $u$ is $s_i$ and the embedding vector is $\mathbf{E}^{u}_i$, the forward propagation process in \textit{embedding input} part is: $\widehat{\mathbf{E}}^{u}_{i+1} = \mathbf{W}_{i i+1}\mathbf{E}^{u}_i + \mathbf{b}_i,..,\widehat{\mathbf{E}}^{u}_n = \mathbf{W}_{n-1 n}\widehat{\mathbf{E}}^{u}_{n-1} + \mathbf{b}_{n-1} $.
%\begin{center}
%\begin{equation}
%\begin{aligned}
%\hat{E_{i+1}} &= W_{i i+1}E_i + b_i\\
%&...\\
%\hat{E_n} &= W_{n-1 n}\hat{E_{n-1}} + b_{n-1}
%\end{aligned}
%\end{equation}
%\end{center}
The embedding vector will be transformed into the $s_n$-dimensional space $\mathbb{R}^{s_n}$. Then, an additional batch normalization with Tanh activation is necessary to tackle the magnitude differences between inner-batch transformed embeddings $\widehat{\mathbf{E}}^{u}_n$ if processing a mini-batch:$\widehat{\mathbf{E}}^{u}_n = tanh \left(\frac{\widehat{\mathbf{E}}^{u}_n - \mu_B}{\sqrt{(\sigma_B^2)^2 + \epsilon}} \right)$,
%\begin{equation}
%\begin{aligned}
%\widehat{\mathbf{E}}^{u}_n = tanh \left(\frac{\widehat{\mathbf{E}}^{u}_n - %\mu_B}{\sqrt{(\sigma_B^2)^2 + \epsilon}} \right),
%\end{aligned}
%\end{equation}
where $\mu_B$ is the mini-batch mean and $\sigma_B^2$ is the mini-batch variance. The batch size can be set as $1$ when inferring a single sample. After executing a similar transformation on item $i$, we obtain the transformed user embedding $\widehat{\mathbf{E}}^u_n$ and item embedding $\widehat{\mathbf{E}}^i_n$ with the same dimension $s_n$. 
The following \textit{representation learning} part is a sequence of BatchNorm, Linear, and Tanh activation layers: $\mathbf{h}_1 = BatchNorm(cat(\widehat{\mathbf{E}}^u_n, \widehat{\mathbf{E}}^i_n)),
\mathbf{h}_2 = BatchNorm(Linear(\mathbf{h}_1)),
\hat{\mathbf{y}}_{ui} = Linear(Tanh(\mathbf{h}_2)).$
%\begin{equation}
%\begin{aligned}
%\mathbf{h}_1 &= BatchNorm(cat(\widehat{\mathbf{E}}^u_n, \widehat{\mathbf{E}}^i_n))\\
%\mathbf{h}_2 &= BatchNorm(Linear(\mathbf{h}_1)) \\
%\hat{\mathbf{y}}_{ui} &= Linear(Tanh(\mathbf{h}_2)). \\
%\end{aligned}
%\end{equation}
\subsubsection{Embedding Warm Initialization}
When receiving the increasing embedding size command from $\pi_{SE}$, there are two intuitive ways to initialize the embedding vector with the new embedding size: 1) zero initialization/random initialization, and 2) initialization with the information from previous embedding vectors. We take the second type of initialization and name it as \textit{embedding warm initialization} (EWI). We perform a linear transformation sharing the parameters with above on the previous embedding $\textbf{E}_i \in \mathbb{R}^{s_i}$ and obtain $\textbf{E}_{i+1}$ in the $s_{i+1}$-dimensional space $\mathbb{R}^{s_{i+1}}$: $\mathbf{E}_{i+1} = \mathbf{W}_{i i+1}\mathbf{E}_i + \mathbf{b}_i$. 
%\begin{equation}
%    \mathbf{E}_{i+1} = \mathbf{W}_{i i+1}\mathbf{E}_i + \mathbf{b}_i.
%\end{equation}

After the embedding warm initialization, the model inference starts from $\textbf{E}_{i+1}$ and follows the forward propagation introduced in Section \ref{subsection:mi}. 

\section{Experiments}
\label{sec:exp}
 In this section, to comprehensively demonstrate the effectiveness of our method, we mainly focus on the following questions:

\begin{itemize}[leftmargin=*]
% \vspace{-3mm}

\item \textbf{RQ1:} Does our method achieve better recommendation accuracy than the state-of-art methods along the timeline?

\item \textbf{RQ2:} Does our method get sublinear regret on selecting embedding sizes, outperforming previous methods?

\item \textbf{RQ3:} Whether our method consumes less computer memory compared with baseline methods?

\item \textbf{RQ4:} Whether our method is more time-efficient than baselines for streaming recommendation?

\item \textbf{RQ5:} Whether our method is applicable to different base recommendation models, like matrix factorization-based model and distance-based model?

\item \textbf{RQ6:} Whether \textit{Embedding Warm Initialization} technique contributes to the model performance improvement?

\end{itemize}

\subsection{Experiment Setting}
\subsubsection{\textbf{Datasets}}
We evaluate our method on four public recommender system datasets. The data and code will be released soon.%\footnote{The data and code will be released soon.}.
\begin{itemize}[leftmargin=*]
    \item \textbf{ml-20m~\citep{harper2015movielens}:} This dataset describes 5-star rating and free-text tagging activity from MovieLens, a movie recommendation service. It contains 20,000,263 ratings created by 138,493 users over 27,278 movies between January 09, 1995 and March 31, 2015.% Note that the MovieID in this dataset is not consecutive integers. The maximum MovieID is 131170 which is far beyond the total number of the movies. 
    \item \textbf{ml-latest~\citep{harper2015movielens}:} This is a  recently released dataset from MovieLens and contains 27,753,444 ratings by 283,228 users over 58,098 movies between January 09, 1995 and September 26, 2018. %The maximum MovieID is 187595.
    \item \textbf{Amazon-Books~\citep{he2016ups}:} This dataset contains book reviews from Amazon, including 22,507,154 ratings spanning May 1996 - July 2014. We download the ratings-only dataset and preprocess it like~\citep{sun2021hgcf}. In the filtered dataset, each user has reviewed at least 20 books and each book has been reviewed by at least 20 users. 
    \item \textbf{Amazon-CDs~\citep{he2016ups}:} This is a CD review dataset also from the website above, including 3,749,003 ratings covering the same time span. A similar preprocessing operation is executed and the filtering threshold is set as 10.
    
\end{itemize}

\subsubsection{\textbf{Tasks}}
We adopt the top-$k$ recommendation and rating score prediction tasks to evaluate the effectiveness of our method.
%Similar to previous works~\citep{liu2018darts, zhaok2021autoemb, liu2020automated}, we utilize the following two rating score prediction tasks as the benchmark:
\begin{itemize}[leftmargin=*]
    \item \textbf{Top-$k$ Recommendation:} This is one of the most common recommendation tasks to evaluate the model's ability on inferring users' intentions. In detail, the model needs to recommend a list of items with the length $k$ to each user according to their historical interaction records. The accuracy is measured with metrics Recall@$k$ and NDCG@$k$. Recall@$k$ indicates what percentage of a user's rated items can emerge in the list. NDCG@$k$ is the normalized discounted cumulative gain at $k$, which takes the position of correctly recommended items into consideration. Here, we take $k$ to be 10.
    %On each dataset, the rating scores for items vary from 0 to 5.0 in 0.5 intervals. When preprocessing the raw data, we set the rating scores greater than the threshold 3.5 to 1.0 and others to 0.0. In the model test phase, if the prediction score is greater than 0.5, this prediction will be regarded as a success, otherwise a failure. The model performance is measured with classification accuracy and mean-squared-error loss. In the reality, this task can be understood as predicting if user will purchase an item.
    \item \textbf{Rating Score Prediction:} Similar to previous works~\citep{liu2018darts, zhaok2021autoemb, liu2020automated}, we also utilize the following two rating score prediction subtasks as the benchmark: \textbf{binary classification} and \textbf{multiclass classification}. The former can be understood as predicting if a user likes an item. And the latter can be used to estimate the discretized interest degree of users. For binary classification task, when preprocessing the raw data, we set the rating scores greater than the threshold 3.5 to 1.0 and others to 0.0. The model performance is measured with classification accuracy and mean-squared-error loss~\citep{liu2020automated,zhaok2021autoemb}. For multiclass classification task, we regard the 5-star rating scores as 5 classes. The model performance is measured with classification accuracy and cross-entropy loss~\citep{liu2020automated,zhaok2021autoemb}.
    
    %When preprocessing the data on each dataset, we round down the ratings by subtracting 0.5 first and the label will vary from 0 to 4 with the interval of 1. In the model test phase, the recommender model needs to output a 5-dimensional vector representing the score for each class. If the class with the maximum score is same as the label, this classification will be regarded as a success, otherwise a failure. The model performance is measured with classification accuracy and cross-entropy loss. In the reality, this classification task can be used to estimate the discretized interest degree of users.
% 2x4 subfigures    
\end{itemize}
\vspace{-0.2cm}

\begin{table*}[]
\caption{The top-$k$ recommendation performance on all four datasets. Our results are statistically significant (t-test, $ p <= 0.01 $).}
\begin{tabular}{|c|cc|cc|cccc|}
\hline
\multirow{2}{*}{Methods} & \multicolumn{2}{c|}{ml-20m}                            & \multicolumn{2}{c|}{ml-latest}                         & \multicolumn{2}{c|}{Amazon-Books}                                           & \multicolumn{2}{c|}{Amazon-CDs}                        \\ \cline{2-9} 
                         & \multicolumn{1}{c|}{Recall@10}       & NDCG@10         & \multicolumn{1}{c|}{Recall@10}       & NDCG@10         & \multicolumn{1}{c|}{Recall@10}       & \multicolumn{1}{c|}{NDCG@10}         & \multicolumn{1}{c|}{Recall@10}       & NDCG@10         \\ \hline
Fixed-128                & \multicolumn{1}{c|}{0.0774}          & 0.0785          & \multicolumn{1}{c|}{0.0779}          & 0.0783          & \multicolumn{1}{c|}{0.0568}          & \multicolumn{1}{c|}{0.0404}          & \multicolumn{1}{c|}{0.0759}          & 0.0402          \\ \hline
Fixed-222                & \multicolumn{1}{c|}{0.0769}          & 0.0779          & \multicolumn{1}{c|}{0.0785}          & 0.0791          & \multicolumn{1}{c|}{0.0573}          & \multicolumn{1}{c|}{0.0416}          & \multicolumn{1}{c|}{0.0742}          & 0.0397          \\ \hline
DARTS                    & \multicolumn{1}{c|}{0.0786}          & 0.0791          & \multicolumn{1}{c|}{0.0784}          & 0.0795          & \multicolumn{1}{c|}{0.0597}          & \multicolumn{1}{c|}{0.0435}          & \multicolumn{1}{c|}{0.0778}          & 0.0432          \\ \hline
AutoEmb                  & \multicolumn{1}{c|}{0.0771}          & 0.0782          & \multicolumn{1}{c|}{0.0783}          & 0.0792          & \multicolumn{1}{c|}{0.0571}          & \multicolumn{1}{c|}{0.0412}          & \multicolumn{1}{c|}{0.0780}          & 0.0429          \\ \hline
ESAPN                    & \multicolumn{1}{c|}{0.0831}          & 0.0842          & \multicolumn{1}{c|}{0.0825}          & 0.0837          & \multicolumn{1}{c|}{0.0632}          & \multicolumn{1}{c|}{0.0475}          & \multicolumn{1}{c|}{0.0816}          & 0.0468          \\ \hline
DESS-FRE (w/o EWI)       & \multicolumn{1}{c|}{\textbf{0.0858}} & \textbf{0.0867} & \multicolumn{1}{c|}{\textbf{0.0866}} & \textbf{0.0869} & \multicolumn{1}{c|}{\textbf{0.0658}} & \multicolumn{1}{c|}{\textbf{0.0494}} & \multicolumn{1}{c|}{\textbf{0.0843}} & \textbf{0.0487} \\ \hline
DESS-FRE                 & \multicolumn{1}{c|}{\textbf{0.0866}} & \textbf{0.0876} & \multicolumn{1}{c|}{\textbf{0.0874}} & \textbf{0.0879} & \multicolumn{1}{c|}{\textbf{0.0672}} & \multicolumn{1}{c|}{\textbf{0.0507}} & \multicolumn{1}{c|}{\textbf{0.0852}} & \textbf{0.0493} \\ \hline
DESS-CV (w/o EWI)        & \multicolumn{1}{c|}{\textbf{0.0875}} & \textbf{0.0888} & \multicolumn{1}{c|}{\textbf{0.0879}} & \textbf{0.0887} & \multicolumn{4}{c|}{\multirow{2}{*}{N/A}}                                                                                            \\ \cline{1-5}
DESS-CV                  & \multicolumn{1}{c|}{\textbf{0.0898}} & \textbf{0.0891} & \multicolumn{1}{c|}{\textbf{0.0884}} & \textbf{0.0895} & \multicolumn{4}{c|}{}                                                                                                                \\ \hline
\end{tabular}
\label{table:top-k}
\end{table*}

\begin{table*}[ht]
\caption{The performance of both rating score binary classification and rating score multiclass classification tasks on ml-20m dataset and ml-latest dataset. Our results are statistically significant (t-test, $ p <= 0.01 $).}
\begin{tabular}{|c|cccc|cccc|l}
\cline{1-9}
\multirow{3}{*}{Methods} & \multicolumn{4}{c|}{ml-20m}                                                                                                            & \multicolumn{4}{c|}{ml-latest}                                                                                                                           & \multicolumn{1}{c}{} \\ \cline{2-9}
                        & \multicolumn{2}{c|}{Binary Classification Task}                                         & \multicolumn{2}{c|}{Multi Classification Task}                & \multicolumn{2}{c|}{Binary Classification Task}                                         & \multicolumn{2}{c|}{Multiclass Classification Task}                                  &                      \\ \cline{2-9}
                         & \multicolumn{1}{c|}{Accuracy} & \multicolumn{1}{c|}{Loss}    & \multicolumn{1}{c|}{Accuracy} & Loss    & \multicolumn{1}{c|}{ Accuracy} & \multicolumn{1}{c|}{Loss}    & \multicolumn{1}{l|}{Accuracy} & \multicolumn{1}{c|}{Loss} &                      \\ \cline{1-9}
Fixed-128                & \multicolumn{1}{c|}{70.93\%}          & \multicolumn{1}{c|}{0.1904}          & \multicolumn{1}{c|}{47.88\%}          & 1.1870          & \multicolumn{1}{c|}{70.98\%}          & \multicolumn{1}{c|}{0.1900}          & \multicolumn{1}{c|}{48.28\%}          & 1.1804                            &                      \\ \cline{1-9}
Fixed-222                & \multicolumn{1}{c|}{71.09\%}          & \multicolumn{1}{c|}{0.1896}          & \multicolumn{1}{c|}{48.19\%}          & 1.1799          & \multicolumn{1}{c|}{71.13\%}          & \multicolumn{1}{c|}{0.1892}          & \multicolumn{1}{c|}{48.62\%}          & 1.1727                            &                      \\ \cline{1-9}
DARTS                    & \multicolumn{1}{c|}{71.07\%}          & \multicolumn{1}{c|}{0.1897}          & \multicolumn{1}{c|}{48.24\%}          & 1.1781          & \multicolumn{1}{c|}{71.12\%}          & \multicolumn{1}{c|}{0.1892}          & \multicolumn{1}{c|}{48.73\%}          & 1.1702                            &                      \\ \cline{1-9}
AutoEmb                  & \multicolumn{1}{c|}{70.54\%}          & \multicolumn{1}{c|}{0.1917}          & \multicolumn{1}{c|}{47.99\%}          & 1.1820          & \multicolumn{1}{c|}{71.00\%}          & \multicolumn{1}{c|}{0.1892}          & \multicolumn{1}{c|}{48.61\%}          & 1.1718                            &                      \\ \cline{1-9}
ESAPN                    & \multicolumn{1}{c|}{71.62\%}          & \multicolumn{1}{c|}{0.1861}          & \multicolumn{1}{c|}{49.24\%}          & 1.1539          & \multicolumn{1}{c|}{71.40\%}          & \multicolumn{1}{c|}{0.1870}          & \multicolumn{1}{c|}{49.52\%}          & 1.1510                            &                      \\ \cline{1-9}
DESS-FRE (w/o EWI)                & \multicolumn{1}{c|}{\textbf{71.89\%}} & \multicolumn{1}{c|}{\textbf{0.1843}} & \multicolumn{1}{c|}{\textbf{49.55\%}} & \textbf{1.1463} & \multicolumn{1}{c|}{\textbf{71.77\%}} & \multicolumn{1}{c|}{\textbf{0.1849}} & \multicolumn{1}{c|}{\textbf{49.89\%}} & \textbf{1.1435}                   &                      \\ \cline{1-9}
DESS-FRE                & \multicolumn{1}{c|}{\textbf{71.93\%}} & \multicolumn{1}{c|}{\textbf{0.1837}} & \multicolumn{1}{c|}{\textbf{49.67\%}} & \textbf{1.1438} & \multicolumn{1}{c|}{\textbf{71.98\%}} & \multicolumn{1}{c|}{\textbf{0.1838}} & \multicolumn{1}{c|}{\textbf{50.05\%}} & \textbf{1.1383}                   &                      \\ \cline{1-9}
DESS-CV (w/o EWI)                 & \multicolumn{1}{c|}{\textbf{72.29\%}} & \multicolumn{1}{c|}{\textbf{0.1823}} & \multicolumn{1}{c|}{\textbf{49.98\%}} & \textbf{1.1358} & \multicolumn{1}{c|}{\textbf{72.35\%}} & \multicolumn{1}{c|}{\textbf{0.1819}} & \multicolumn{1}{c|}{\textbf{50.24\%}} & \textbf{1.1327}                   &                      \\ \cline{1-9}
DESS-CV                 & \multicolumn{1}{c|}{\textbf{73.28\%}} & \multicolumn{1}{c|}{\textbf{0.1768}} & \multicolumn{1}{c|}{\textbf{51.19\%}} & \textbf{ 1.1103} & \multicolumn{1}{c|}{\textbf{73.07\%}} &\multicolumn{1}{c|}{\textbf{0.1779}} & \multicolumn{1}{c|}{\textbf{50.73\%}} & \textbf{1.1216}                   &                      \\ \cline{1-9}
\end{tabular}
\label{table:movielens}
\end{table*}        

\begin{table*}[ht]
\caption{The performance of both rating score binary classification and rating score multiclass classification tasks on Amazon-Books dataset and Amazon-CDs dataset. Our results are statistically significant (t-test, $ p <= 0.01 $).}
\begin{tabular}{|c|cccc|cccc|l}
\cline{1-9}
\multirow{3}{*}{Methods} & \multicolumn{4}{c|}{Amazon-Books}                                                                                                      & \multicolumn{4}{c|}{Amazon-CDs}                                                                                                                      & \multicolumn{1}{c}{} \\ \cline{2-9}
                         & \multicolumn{2}{c|}{Binary Classification Task}                                         & \multicolumn{2}{c|}{Multi Classification Task}                & \multicolumn{2}{c|}{Binary Classification Task}                                         & \multicolumn{2}{c|}{Multiclass Classification Task}                              &                      \\ \cline{2-9}
                         & \multicolumn{1}{c|}{Accuracy}     & \multicolumn{1}{c|}{Loss}        & \multicolumn{1}{c|}{Accuracy}     & Loss        & \multicolumn{1}{c|}{Accuracy}     & \multicolumn{1}{c|}{Loss}        & \multicolumn{1}{l|}{Accuracy}     & \multicolumn{1}{c|}{Loss} &                      \\ \cline{1-9}
Fixed-128                & \multicolumn{1}{c|}{80.75\%}          & \multicolumn{1}{c|}{0.1459}          & \multicolumn{1}{c|}{54.31\%}          & 1.1225          & \multicolumn{1}{c|}{80.28\%}          & \multicolumn{1}{c|}{0.1545}          & \multicolumn{1}{c|}{57.17\%}          & 1.1638                        &                      \\ \cline{1-9}
Fixed-222                & \multicolumn{1}{c|}{80.87\%}          & \multicolumn{1}{c|}{0.1433}          & \multicolumn{1}{c|}{54.85\%}          & 1.1026          & \multicolumn{1}{c|}{80.09\%}          & \multicolumn{1}{c|}{0.1540}          & \multicolumn{1}{c|}{57.24\%}          & 1.1536                        &                      \\ \cline{1-9}
DARTS                    & \multicolumn{1}{c|}{81.02\%}          & \multicolumn{1}{c|}{0.1415}          & \multicolumn{1}{c|}{55.22\%}          & 1.0895          & \multicolumn{1}{c|}{80.50\%}          & \multicolumn{1}{c|}{0.1498}          & \multicolumn{1}{c|}{57.74\%}          & 1.1305                        &                      \\ \cline{1-9}
AutoEmb                  & \multicolumn{1}{c|}{80.60\%}          & \multicolumn{1}{c|}{0.1456}          & \multicolumn{1}{c|}{54.77\%}          & 1.0910          & \multicolumn{1}{c|}{80.48\%}          & \multicolumn{1}{c|}{0.1504}          & \multicolumn{1}{c|}{57.89\%}          & 1.1232                        &                      \\ \cline{1-9}
ESAPN                    & \multicolumn{1}{c|}{81.51\%}          & \multicolumn{1}{c|}{0.1357}          & \multicolumn{1}{c|}{56.84\%}          & 1.0437          & \multicolumn{1}{c|}{81.44\%}          & \multicolumn{1}{c|}{0.1399}          & \multicolumn{1}{c|}{59.50\%}          & 1.0616                        &                      \\ \cline{1-9}
DESS-FRE (w/o EWI)                 & \multicolumn{1}{c|}{\textbf{81.62\%}} & \multicolumn{1}{c|}{\textbf{0.1342}} & \multicolumn{1}{c|}{\textbf{57.14\%}} & \textbf{1.0372} & \multicolumn{1}{c|}{\textbf{81.82\%}} & \multicolumn{1}{c|}{\textbf{0.1367}} & \multicolumn{1}{c|}{\textbf{60.02\%}} & \textbf{1.0453}               &                      \\ \cline{1-9}
DESS-FRE                 & \multicolumn{1}{c|}{\textbf{81.76\%}} & \multicolumn{1}{c|}{\textbf{0.1336}} & \multicolumn{1}{c|}{\textbf{57.35\%}} & \textbf{1.0329} & \multicolumn{1}{c|}{\textbf{81.93\%}} & \multicolumn{1}{c|}{\textbf{0.1361}} & \multicolumn{1}{c|}{\textbf{60.15\%}} & \textbf{1.0491}               &                      \\ \cline{1-9}
\end{tabular}
\label{table:amazon}
\end{table*}

\subsubsection{\textbf{Baselines}} 
Following methods serve as the baselines:
\begin{itemize}[leftmargin=*]
\item \textbf{Fixed}: The base Neural Collaborative Filtering~\citep{he2017neural} model, where the embedding sizes for users and items are both identical and fixed. To fairly compare experimental results, , we set the embedding sizes to 128 and 222, regarding the two versions \textbf{Fixed-128} and \textbf{Fixed-222} of the model, respectively. 
%\item SMAC
%\item Nelder-Mead Approach
\item \textbf{DARTS}~\citep{liu2018darts}: A type of \textit{soft-selection} algorithm developed from the neural architecture search. The weight vectors regarding embedding sizes are trained directly with gradient backpropagation.%The input of Neural CF layers is the weighted summation of transformed vectors corresponding to each embedding size. The weight vectors are updated with the backpropagation of the validation loss,  during which the recommendation model parameters will not be influenced.
\item \textbf{AutoEmb}~\citep{zhaok2021autoemb}: Another type of \textit{soft-selection} algorithm similar to DARTS. However, the weight vectors are the outputs of controller neural networks independent of the recommendation model.% The input of controller includes the frequency of the user/item and the contextual information such as the previous hyperparameters and the corresponding loss the user/item obtains.
\item \textbf{ESAPN}~\citep{liu2020automated}: A type of \textit{hard-selection} algorithm using the REINFORCE algorithm~\citep{williams1992simple} as the controller to dynamically choose the suitable embedding size for corresponding users and items. %According to the original paper, the controller inputs include the frequencies and current embedding sizes.

\end{itemize}
\vspace{-0.1cm}

Due to the fact that Amazon datasets cannot provide the raw item feature vectors, we run both \textbf{DESS-CV} (with $(FRE^u_l, IND^u_l)$ and $(FRE^i_l, POD^i_l)$ as embedding size search policies input) and \textbf{DESS-FRE} (with only historical frequency as search policies input) on ml-20m and ml-latest datasets while only \textbf{DESS-FRE} on Amazon-Books and Amazon-CDs datasets. The comparison between DESS-CV and DESS-FRE can be regarded as the ablation study to the embedding size indicators proposed in Section~\ref{section:infana}. All the results below are the average  of five trials with different random seeds.
\vspace{-0.3cm}

\subsection{Results and Analysis}
\subsubsection{\textbf{Recommendation Accuracy (RQ1)}} In Tables~\ref{table:top-k}, ~\ref{table:movielens} and~\ref{table:amazon}, we report the average performance of recommendation models on test data sequence $\{D^{te}_{1},...,D^{te}_{T}\}$ of each task. First, we observe that DESS performs significantly better than the \textit{fixed} methods and the \textit{soft-selection} methods. This proves that \textit{hard-selection}, generally speaking, is a more effective way to search embedding sizes. Second, we notice that our DESS-FRE and DESS-CV achieve better results compared with the state-of-the-art \textit{hard-selection} algorithm ESAPN on all tasks and datasets. This demonstrates the effectiveness of DESS algorithm in improving the streaming recommendation. Third, by comparing the results of DESS-FRE with ESAPN and DESS-CV with DESS-FRE, respectively, we can find that the non-stationary LinUCB-based search policy $\pi_{SE}$ and the two indicators ($IND$ and $POD$) both contribute to the performance gains.

\subsubsection{\textbf{Embedding Size Selection Regret (RQ2)}} 
\begin{figure*}[ht]
    \centering
    \includegraphics[width=0.97\textwidth]{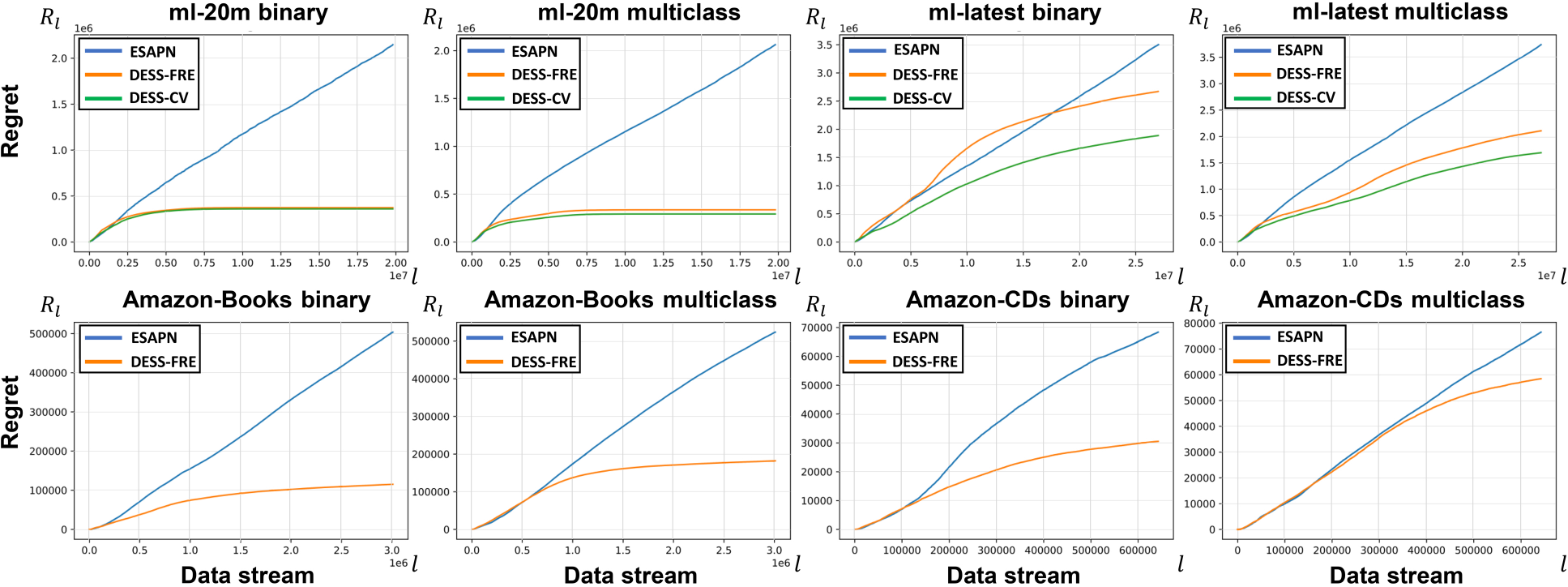}
    \caption{The regret curves of three \textit{hard-selection} methods: ESAPN, DESS-FRE, DESS-CV across binary classification and multiclass classification tasks. DESS-CV cannot be evaluated on the Amazon datasets due to the lack of raw item features.
    }
    \label{fig:regret_curves}
    \vspace{-0.3cm}
\end{figure*}
%\begin{figure}[h]
%    \centering
%    \includegraphics[width=0.48\textwidth]{Figures/histogram of regret.pdf}
%    \caption{The regret and pseudo-regret of binary classification and multiclass classification tasks on four datasets. ``BC'' denotes  binary classification, ``MC'' denotes multiclass classification. ``R'' indicates regret and ``Pr'' indicates pseudo-regret.
%    }
%    \label{fig:regret_histogram}
%    \vspace{-0.3cm}
%\end{figure}
We report the regret in terms of users due to the space limitation. The item side has a similar trend. Fixed embedding size methods and \textit{soft-selection} methods have no regret because they actually use the embeddings of all sizes at the same time. We illustrate the regret curves of two rating score prediction tasks on each dataset in Figure~\ref{fig:regret_curves}. %The overall regret and pseudo-regret are also shown in Figure~\ref{fig:regret_histogram}.  
First, the observed decline in regret connects the aforementioned improvement in accuracy, justifying the advantage of modeling the dynamic embedding size search as bandits. Second, the regret can be reduced several times to dozens of times compared with ESAPN. The regret of DESS-CV is a bit lower than that of DESS-FRE. These demonstrate the effectiveness of $\pi_{SE}$ and also the two indicators. Third, from the regret curves, we observe the sublinear increase phenomena of DESS-FRE and DESS-CV, which is also in line with the regret upper bound guarantee given in Section~\ref{section:theory}.

\begin{figure}
    \centering
    \includegraphics[width=0.45\textwidth]{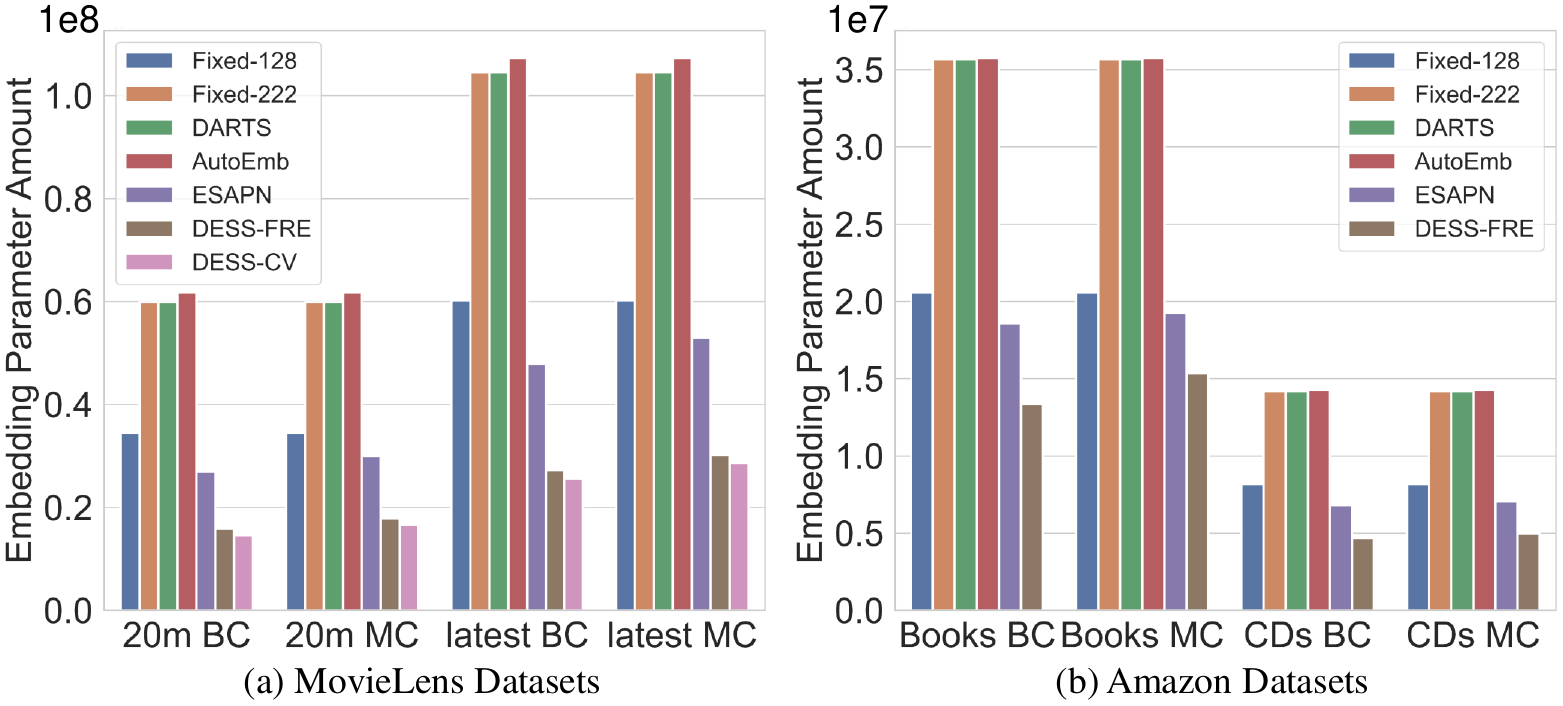}
    \caption{The parameter comparison among different methods. In the table, ``BC'' refers to the binary classification task and ``MC' denotes the multiclass classification task.
    }
    \label{fig:emb para}
    \vspace{-0.3cm}
\end{figure}

\subsubsection{\textbf{Model Memory Cost (RQ3)}} In Figure~\ref{fig:emb para}, we calculate the average number of embedding parameters for each algorithm on each task. From this figure, we observe that the embedding parameter quantity of DESS-FRE is much less than that of other four methods. As for ml-20m, the memory consumption of Fixed-128 and two \textit{soft selection} approaches are 1.93 times, 3.35 times, and 3.45 times of DESS-FRE. In the recommendation tasks on ml-latest dataset, our method only consumes $50.2\%$, $28.9\%$, $28.9\%$ and $28.2\%$ memory compared with the Fixed-128, Fixed-222, DARTS, and AutoEmb, respectively. Moreover, the embedding memory cost brought by DESS-CV is even less than that of DESS-FRE. Similar memory overhead savings can also be observed on Amazon datasets. These all certify that our proposed methods can reduce the memory cost effectively and contribute to the more efficient algorithm DESS.

\subsubsection{\textbf{Time Efficiency Analysis (RQ4)}}
\begin{figure}[ht]
    \centering
    \includegraphics[width=0.45\textwidth]{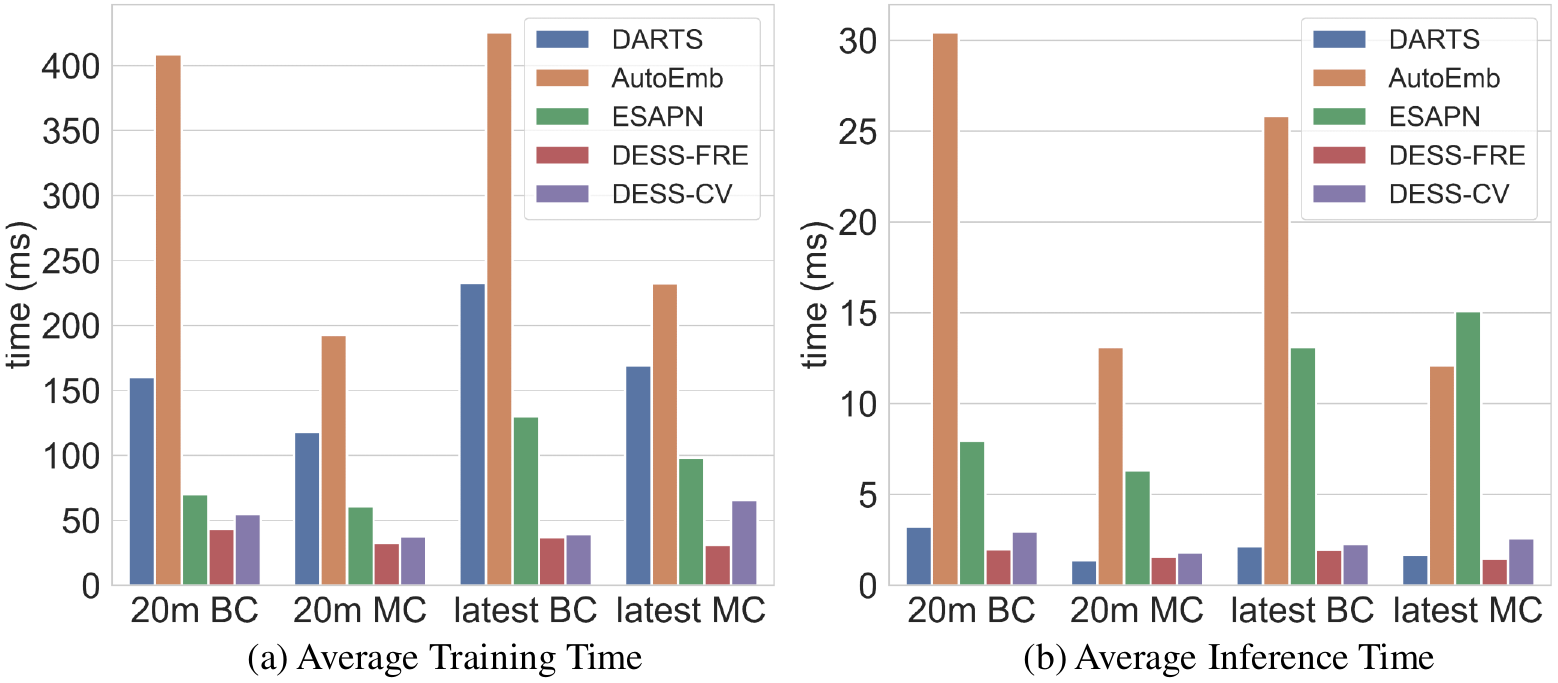}
    \caption{The average training time and inference time of different embedding size search methods in two rating score prediction tasks on ml-latest dataset and ml-20m dataset. 
    }
    \label{fig:time analysis}
    \vspace{-0.3cm}
\end{figure}
%\begin{figure}
%	\centering
%	\subfigure{
%		\includegraphics[width=0.9\linewidth]{Figures/average training time.png}
%    }
%	\subfigure{
%	\includegraphics[width=0.9\linewidth]{Figures/average inference time.png}
%	}
  %   \subfigure[{existing item graph}]{
		% \includegraphics[width=0.3\linewidth]{figures/Instruments_ILAD-NDCG.png}
  %   }
%    \caption{The average training time and inference time of two rating score prediction tasks on ml-latest dataset and ml-20m dataset. ``BC'' denotes binary classification, ``MC'' denotes multiclass classification. ``latest'' indicates ml-latest dataset and ``20m'' indicates ml-20m dataset..}
%    \label{fig:time}
%\end{figure}
Time efficiency is also of great significance when deploying streaming recommendation models in real world. In the experiments, we count the average training time and average inference time of our method and embedding size-changeable baselines for the above two classification tasks on ml-latest and ml-20m datasets. The results are shown in Figure~\ref{fig:time analysis}. From the figure, we can observe that the average training time of our \textbf{DESS} algorithms is much less that of other \textit{soft-selection} and \textit{hard-selection} methods. More precisely, the training time of DESS-FRE is no more than $20\% \sim 30\%$ of Darts and is even no more than $10\%$ of AutoEmb. We can also observe a similar trend in average inference time histogram. Thus, we can speculate that our DESS algorithm holds obvious time efficiency advantage compared with previous methods. This is actually because that the bandit inherently has more lighter model and faster decision-making process compared with deep neural work-based and reinforcement learning-based embedding size selection policies.  Besides, we can find that both the average training time and average inference time of DESS-FRE are less than DESS-CV to some extent. This is due to the lower input dimension and the smaller linear matrix in the bandit's reward model, which finally lead to much faster computation.

\subsubsection{\textbf{Method Applicability Analysis (RQ5)}}
\label{section:furana}
We also explore the effects of our method when choosing the matrix factorization-based model and distance-based model as the base streaming recommendation models, respectively. Different dynamic embedding size search methods are evaluated with the rating score binary classification task on ml-20m and ml-latest datasets. From the accuracy reported in Table~\ref{table:furtheranalysis}, it can be first noticed that the recommendation models with fixed embedding sizes (Fixed-128 and Fixed-222) are much worse than that with dynamic embedding sizes. This also confirms the necessity of the dynamic embedding size search in the streaming recommendation. Second, we observe that DESS-CV outperforms all the baselines regardless of datasets and recommendation models.  Even the DESS-FRE without the support from $IND$ and $POD$ is still superior to all the previous methods in most cases. Such experimental results demonstrate that our algorithm can be a general approach towards more effective dynamic embedding size search in streaming recommendation.

\begin{table}[]
\caption{The binary classification accuracy when adopting matrix factorization-based model and distance-based model. Our results are statistically significant (t-test, $ p <= 0.01 $).}
\begin{tabular}{|c|cc|cc|}
\hline
\multirow{2}{*}{Methods} & \multicolumn{2}{c|}{Matrix Factorization-based}               & \multicolumn{2}{c|}{Distance-based}                           \\ \cline{2-5} 
                         & \multicolumn{1}{c|}{ml-20m} & ml-latest & \multicolumn{1}{c|}{ml-20m} & ml-latest  \\ \hline
Fixed-128                & \multicolumn{1}{c|}{50.03\%}                  &       50.08\%               & \multicolumn{1}{c|}{54.87\%}                  &       55.36\%               \\ \hline
Fixed-222                & \multicolumn{1}{c|}{50.07\%}                  &         50.04\%             & \multicolumn{1}{c|}{52.02\%}                  &        52.59\%              \\ \hline
DARTS                    & \multicolumn{1}{c|}{70.74\%}                  &           70.92\%           & \multicolumn{1}{c|}{70.17\%}                  &         70.17\%             \\ \hline
AutoEmb                  & \multicolumn{1}{c|}{69.77\%}                  &         70.32\%             & \multicolumn{1}{c|}{70.30\%}                  &      70.44\%                \\ \hline
ESAPN                    & \multicolumn{1}{c|}{70.21\%}                  &      71.86\%                & \multicolumn{1}{c|}{\textbf{71.55\%}}                  &      71.03\%                \\ \hline
DESS-FRE                 & \multicolumn{1}{c|}{\textbf{72.75\%}}                  &         \textbf{72.61\%}             & \multicolumn{1}{c|}{71.46\%}                  &        \textbf{71.84\%}              \\ \hline
DESS-CV                  & \multicolumn{1}{c|}{\textbf{73.84\%}}                  &          \textbf{73.25\%}           & \multicolumn{1}{c|}{\textbf{73.02\%}}                  &           \textbf{72.49\%}            \\ \hline
\end{tabular}
\label{table:furtheranalysis}
\end{table}
%\vspace{2mm}

\subsubsection{\textbf{Ablation Study (RQ6)}}
To validate the effectiveness of \textit{Embedding Warm Initialization} technique proposed in Section~\ref{section:esann}, we also conduct the experiments of DESS-FRE (w/o EWI) and DESS-CV (w/o EWI) on four datasets, whose results are shown in Tables~\ref{table:top-k},~\ref{table:movielens}, and~\ref{table:amazon}. We can observe that, DESS-FRE and DESS-CV both achieve stable performance gain over DESS-FRE (w/o EWI) and DESS-CV (w/o EWI), respectively, especially on ml-20m and ml-latest datasets. The limited improvement on Amazon datasets may be due to the bottleneck of the base recommender model itself. These results demonstrate that initializing the embeddings with previous information via a simple linear transformation can effectively benefit the recommendation performance along the timeline.

\section{Conclusion}
%Traditional deep recommendation models often use the embedding vectors with identical and fixed dimensions to convert discrete input to continuous ones. However, this kind of design results in huge memory cost and cannot reach optimality in recommendation performance. Some recent works explore to adjust the embedding sizes dynamically in the streaming recommender systems, but still suffer from the unsatisfying performance and unacceptable time delay. 
In this work, we first rethink the streaming model update process and then model the dynamic embedding size as a bandit problem. Based on the embedding size indicator analysis, we provide the DESS algorithm and obtain a sublinear dynamic regret upper bound as the theoretical guarantee. The results of recommendation accuracy, memory cost, and time consumption across various recommendation tasks on four open datasets demonstrate the effectiveness of our method. In the future, we plan to explore the automated tuning methods of other hyperparameters like learning rate in the streaming machine learning.

\begin{acks}
%We thank for the discussion and computation devices from Huawei Noah's Ark Lab, Hong Kong. 
This work was supported by the Start-up Grant (No. 9610564) and the Strategic Research Grant (No. 7005847) of City University of Hong Kong.
\end{acks}

\bibliographystyle{ACM-Reference-Format}
\bibliography{reference.bib}

%\clearpage
%\newpage
\appendix
\section{Notations}
\label{section:notations}
We summarize the main notations used in this paper in Table~\ref{table:notation}.

\begin{table}[h]
\caption{Major notations.}
\begin{tabular}{c l}
\hline
$K$     & \begin{tabular}[c]{@{}l@{}}The number of arms (embedding size candidates)\end{tabular} \\ 
$T$     & \begin{tabular}[c]{@{}l@{}}The total number of time steps \end{tabular} \\ 
$L$               & \begin{tabular}[c]{@{}l@{}}The length of the whole data stream 
% $l$ is the corresponding data  \\point index ranging from 1 to $L$. In the following parts \\ , we  will use variables with $val$ as superscript and $l$ as \\   subscript like $x^{val}_l$ and $a^{val}_l$ which means $l_{th}$ data in the\\  validation data stream. Note that the validation data str-\\   eam is not entirely consistent with the real data stream.
\end{tabular} \\ 
$C$     & \begin{tabular}[c]{@{}l@{}}The set of contexts for non-stationary LinUCB
% In this setting, the contexts are fea- \\ture groups that determine the optimal embedding sizes \\ for the corresponding users or items
\end{tabular} \\ 
$U$     & \begin{tabular}[c]{@{}l@{}} The upper bound of contexts for non-stationary LinUCB
\end{tabular} \\ 
$S$     & \begin{tabular}[c]{@{}l@{}} The upper bound of parameters in reward models of bandit
\end{tabular} \\ 
$\gamma$     & \begin{tabular}[c]{@{}l@{}} The discount factor in non-stationary LinUCB bandit
\end{tabular} \\ 
$d$     & \begin{tabular}[c]{@{}l@{}} The dimension of context vectors for bandit
\end{tabular} \\ 
$\pi^{u}_{SE}$    & \begin{tabular}[c]{@{}l@{}}The dynamic embedding size search policy for users 
\end{tabular} \\ 
$\pi^{i}_{SE}$    & \begin{tabular}[c]{@{}l@{}}The dynamic embedding size search policy for items
\end{tabular} \\ \
$r_l$             & The reward received at $l$-th user-item interaction  ($ l \leq L $) \\ 
$\bm{\theta}_{l, a}$   & \begin{tabular}[c]{@{}l@{}}The updated parameter of the reward model for arm $a$ at \\ $l$-th user-item interaction \end{tabular} \\ 
$\bm{\theta}^*_{l, a}$ & \begin{tabular}[c]{@{}l@{}}The parameter of the oracle reward model for arm $a$ at $l$-th\\ user-item interaction \end{tabular} \\ 
$M_{t}$ & The updated recommendation model at time step $t$ ($ t \leq T $) \\ 
$\mathbf{F}^{i}$ & The raw feature vector for item $i$ \\ \hline
\end{tabular}
\label{table:notation}
\end{table}

\section{Theorem Proof}
\label{section:proof}
\subsection{Proof of Theorem 1}
\label{section: proof of theorem 1}
\begin{proof}
First recall that $a^*_l = \arg max_{a \in \mathcal{A} } \langle \mathbf{x}_l, \bm{\theta}^*_{l,a} \rangle$ and $\mathbf{x}_{l,1} = \mathbf{x}_{l,2} = ... = \mathbf{x}_{l, K} = \mathbf{x}_{l, a^*_l} = \mathbf{x}_l = \mathbf{x}_{(t,j)}(l = t \times |D_t| + j)$,
\begin{equation}
\begin{aligned}
R_L &= \underset{t=1}{\sum^T} \underset{j=1}{\sum^{|D^{val}_t|}} \langle \mathbf{x}_{(t,j)}, \bm{\theta}^*_{(t,j), a^*_{(t, j)}}\rangle - \langle \mathbf{x}_{(t,j)}, \bm{\theta}^*_{(t,j), a_{(t,j)}} \rangle \\
&= \underset{l=1}{\sum^L} \langle \mathbf{x}_l, \bm{\theta}^*_{l, a^*_l} \rangle - \langle \mathbf{x}_l, \bm{\theta}^*_{l, a_l} \rangle\\
&=\underset{l=1}{\sum^L} \langle \mathbf{x}_l, \bm{\theta}^*_{l, a^*_l} \rangle - \langle \mathbf{x}_l, \bm{\theta}_{l, a^*_l} \rangle + \underset{l=1}{\sum^L} \langle \mathbf{x}_l, \bm{\theta}_{l, a^*_l} \rangle - \langle \mathbf{x}_l, \bm{\theta}_{l, a_l} \rangle\\
&+ \underset{l=1}{\sum^L} \langle \mathbf{x}_l, \bm{\theta}_{l, a_l} \rangle - \langle \mathbf{x}_l, \bm{\theta}^*_{l, a_l} \rangle\\
&=\underset{l=1}{\sum^L} \langle \mathbf{x}_l, \bm{\theta}_{l, a^*_l} \rangle - \langle \mathbf{x}_l, \bm{\theta}_{l, a_l} \rangle + \underset{l=1}{\sum^L} \langle \mathbf{x}_l, \bm{\theta}^*_{l, a^*_l} \rangle - \langle \mathbf{x}_l, \bm{\theta}_{l, a^*_l} \rangle\\
&+ \underset{l=1}{\sum^L} \langle \mathbf{x}_l, \bm{\theta}_{l, a_l} \rangle - \langle \mathbf{x}_l, \bm{\theta}^*_{l, a_l} \rangle\\
&\leq \frac{2k}{c} \underset{l=1}{\sum^L} \beta_t[\Vert \mathbf{x}_l \Vert_{V^{-1}_{l,a^*_l}} - \Vert \mathbf{x}_l \Vert_{V^{-1}_{l,a_l}}] +  \underset{l=1}{\sum^L} Er(\mathbf{x}_l, \bm{\theta}_{l,a^*_l})\\
&+ \underset{l=1}{\sum^L} Er(\mathbf{x}_l, \bm{\theta}_{l,a_l}).
\end{aligned}
\end{equation}

Thanks to Lemma \ref{lemma1} and Lemma \ref{lemma2}:
\begin{equation}
\begin{aligned}
R_L &\leq \underbrace{\underset{l=1}{\sum^L}  \frac{4k}{c}\beta_l \Vert \mathbf{x}_{l} \Vert_{V^{-1}_{l,a^*_l}}}_{R^{1}_L} + \underbrace{ \underset{l=1}{\sum^T} \frac{4kU}{c} \sqrt{1 + \frac{U^2}{\lambda(1-\gamma)}}\frac{2kS U^2}{\lambda} \frac{\gamma^D}{1-\gamma}}_{R^{2}_L}\\
& + \underbrace{\underset{l=1}{\sum^L} \frac{4kU}{c} \sqrt{1 + \frac{U^2}{\lambda(1-\gamma)}} \underset{s=l-D}{\sum^{l-1}} k\sqrt{\frac{d}{\lambda(1-\gamma)}} max_{a} \Vert \bm{\theta}^{*}_{s,a} - \bm{\theta}^{*}_{s+1,a}\Vert )}_{R^{3}_L}.
\end{aligned}
\end{equation}

%Next, scale $R^{1}_L$, $R^{2}_L$, and $R^{3}_L$ respectively.
\begin{equation}
\begin{aligned}
R^{1}_L &\leq \underset{l=1}{\sum^L}  \frac{4k}{c}\beta_L \Vert \mathbf{x}_{l} \Vert_{V^{-1}_{l,a^*_l}} \\
& \leq \frac{4k}{c}\beta_L \sqrt{L} \sqrt{\underset{l=1}{\sum^L}\Vert \mathbf{x}_{l} \Vert_{V^{-1}_{l,a^*_l}}} \quad(Cauthy-Schwarz)\\
& \leq \frac{4k}{c}\beta_L \sqrt{2dL\max(1, \frac{U^2}{\lambda})}\sqrt{L\log(\frac{1}{\gamma}) + \log(1+\frac{U^2(1-\gamma^L)}{\lambda d(1-\gamma)})} (Lemma~\ref{lemma2}).
\end{aligned}
\end{equation}

Because $\sqrt{1+\frac{U^2}{\lambda(1-\lambda)}}$ can be upper bounded by $1 + \frac{U}{\sqrt{\lambda(1-\lambda)}}$: \\
\begin{equation}
\begin{aligned}
R^{2}_L &\leq \frac{8 k^2 S U^3 \gamma^D}{c\lambda(1-\gamma)}L + \frac{8 k^2 S U^4 \gamma^D}{c \lambda^{\frac{3}{2}} (1-\gamma)^{\frac{3}{2}}}L \\
R^{3}_L &\leq \frac{4 k^2 UD}{c\sqrt{\lambda}}\sqrt{\frac{d}{1-\gamma}}B_L + \frac{4 k^2 U^2 D}{c\lambda} \frac{\sqrt{d}}{1-\gamma}B_L.
\end{aligned}
\end{equation}

Add such three components together, we have:
\begin{equation}
\begin{aligned}
R_L &\leq R^{1}_L +  R^{2}_L + R^{3}_L \\
& \leq \sqrt{32\max(1,U^2/\lambda)}\frac{k}{c}\beta_L\sqrt{dL}\sqrt{L\log(\frac{1}{\gamma})+\log(1+\frac{U^2(1-\gamma^L)}{\lambda d(1-\gamma)})} + \\
&\frac{8 k^2 S U^3 \gamma^D}{c\lambda(1-\gamma)}L + \frac{8 k^2 S U^4 \gamma^D}{c \lambda^{\frac{3}{2}} (1-\gamma)^{\frac{3}{2}}}L + \frac{4 k^2 UD}{c\sqrt{\lambda}}\sqrt{\frac{d}{1-\gamma}}B_L + \frac{4 k^2 U^2 D}{c\lambda} \frac{\sqrt{d}}{1-\gamma}B_L.
\end{aligned}
\end{equation}

Thus, the theorem \ref{theo1} is proved.
\end{proof}

\subsection{Proof of Corollary 1}
\label{section: proof of corollary 1}
\begin{proof} 
By neglecting the logarithmic term,
\begin{center}
$\beta_L\sqrt{dL}\sqrt{L\log(1/\gamma)} \sim dL\frac{B^{1/5}_L d^{-1/10}}{L^{1/5}} = d^{9/10} B^{1/5}_L L^{4/5}$\\
$\gamma^D L/(1 - \gamma)^{3/2} \sim e^{-\log L} L (\frac{d^{1/5}L^{2/5}}{B_L^{2/5}})^{3/2} = d^{3/10} B^{-3/5}_L L^{3/5}$\\
$\frac{\sqrt{d}}{1-\gamma} D B_L \sim d^{1/2} B_L (\frac{d^{1/5}L^{2/5}}{B_L^{2/5}})^2 = d^{9/10} B^{1/5}_L L^{4/5}$.\\
\end{center}
Thus, with high probability, we have:
\begin{center}
$R_L =\mathcal{O}_{L\rightarrow \infty}(d^{9/10} B^{1/5}_L L^{4/5})$.
\end{center}
\end{proof}

\section{Implementation Details}
\label{section:implementation}
For the fair comparison, we set the embedding size candidates of the last three methods and our DESS as $\{2,4,8,16,64,128\}$, which is also consistent with previous related researches~\citep{liu2018darts, zhaok2021autoemb, liu2020automated}. All the methods share the embedding size adaptive neural network designed in~\ref{section:esann} as the streaming recommendation model. We use the Adam optimizer with an initial learning rate 
as 0.001 and a regularization parameter as 0.001. The hidden layer size of the recommendation model is set to 512. The min-batch size is set as 500. The discount factor $\gamma$ is set as 0.99. The subguassian constant $\sigma$ is set as 3.0. Without specifications, the hyper-parameters are set same as the original paper. We implement our algorithm with PyTorch and test it on the NVIDIA Titan-RTX GPU with 24 GB memory.

%The memory cost comparison among different methods are list below:

\end{document}